\documentclass[%
	abstract=true,	
	11pt,
	dvipsnames,
	numbers=noenddot,
	a4paper,
]{scrartcl}

\usepackage[USenglish,UKenglish]{babel}
\usepackage[utf8]{inputenc}

\usepackage{afterpage}
\usepackage{amsfonts,amsmath,amssymb,amsthm}
\usepackage[blocks,auth-sc-lg,affil-it]{authblk}
\usepackage{appendix}
\usepackage{bm}					
\usepackage{booktabs}					
\usepackage{csquotes}
\usepackage{commath} 								
\usepackage{graphicx}			
\usepackage{longtable}
\usepackage{marvosym}
\usepackage[sc]{mathpazo}			
\usepackage{mathtools}				
\usepackage[
	]{microtype}								
\usepackage{multirow}
\usepackage{nicefrac}
\usepackage{paralist}				
\usepackage{placeins}				
\usepackage{pdflscape}			
\usepackage[%
	onehalfspacing,
	]{setspace}
\usepackage{siunitx}					
\usepackage{subcaption}				
\usepackage[
	disable										
	]{todonotes}

\usepackage{tikz} 
\usetikzlibrary{%
	arrows,
	backgrounds,
	calc,			
	fit,	
	math,
	matrix,
	shapes,			
	positioning}								

\usepackage{pgfplots}
\pgfplotsset{compat=newest}										
\usepgfplotslibrary{patchplots}

\usepackage{xifthen}  			
\usepackage{xspace}

\definecolor{JoyfulColor}{rgb}{0.75,0.0,0.2}
\usepackage[
 	breaklinks=true,
 	urlcolor=JoyfulColor,															
  colorlinks=true,
  linkcolor=JoyfulColor,
  citecolor=JoyfulColor,
]{hyperref}						

\usepackage[%
arxiv=abs,																	
autocite=footnote,													
autolang=hyphen,														
backend=bibtex8,															
backref=true,																
bibstyle=authoryear,												
citestyle=authoryear-comp,									
dashed=false,																
sortcites=false,														
maxnames=2,
maxbibnames=11,															
url=true,doi=true,eprint=true,							
useprefix=true,															
uniquename=false, 													
]{biblatex}

\newcommand*\person[1]{\textsc{#1}}									

\DeclareFieldFormat{pages}{#1}		

\renewbibmacro{in:}{}					

\DefineBibliographyStrings{english}{%
    backrefpage  = {}, 								
    backrefpages = {} 								
}

\newcommand{\red}[1]{{\color{red} #1}}
\newcommand{\blue}[1]{{\color{blue} #1}}

\newcommand{\DarkGreen}[1]{{\color{green!40!black} #1}}

\newcommand{\eg}{\textit{e.g.}\xspace}

\newcommand{\etc}{\textit{etc.}\xspace}

\newcommand{\ie}{\textit{i.e.}\xspace}

\newcommand{\perse}{\textit{per se}\xspace}

\newcommand{\ALWmodel}{\textsc{ALW} model\xspace}
\newcommand{\FWmodel}{\textsc{FW} model\xspace}

\newcommand{\TA}[1]{\ensuremath{\overline{#1}}\xspace}	
\newcommand{\EA}[1]{\ensuremath{\left\langle #1 \right\rangle}\xspace}		
\newcommand{\EV}[1]{\ensuremath{\operatorname{E}\left[ #1 \right]}\xspace}	

\newcommand{\TT}{\ensuremath{\mathbb{T}}\xspace}		

\newcommand{\elabel}[1]{\label{eq:#1}}
\newcommand{\eref}[1]{Eq.~(\ref{eq:#1})}

\newcommand{\flabel}[1]{\label{fig:#1}}
\newcommand{\fref}[1]{Fig.~\ref{fig:#1}}
\newcommand{\Fref}[1]{Figure~\ref{fig:#1}}
\graphicspath{{figs/}} 	

\newcommand{\tlabel}[1]{\label{tab:#1}}
\newcommand{\tref}[1]{Tab.~\ref{tab:#1}}
\newcommand{\Tref}[1]{Table~\ref{tab:#1}}

\newcommand{\seclabel}[1]{\label{sec:#1}}
\newcommand{\sref}[1]{Sec.~\ref{sec:#1}}
\renewcommand{\secref}[1]{Section~\ref{sec:#1}}
\newcommand{\sseclabel}[1]{\label{ssec:#1}}
\newcommand{\ssref}[1]{Subsec.~\ref{ssec:#1}}

\newcommand{\ssseclabel}[1]{\label{sssec:#1}}
\newcommand{\sssref}[1]{Subsubsec.~\ref{sssec:#1}}

\numberwithin{equation}{section}

\makeatletter
\newtheoremstyle{myDefStyle}
   {\topsep}{\topsep}%
   {\itshape}{}%
   {\sffamily\bfseries}{.}%
   { }
   {\thmname{#1}\thmnumber{\@ifnotempty{#1}{ }\@upn{#2}}%
    \thmnote{ {\bfseries(#3)}}}%
\makeatother

\theoremstyle{myDefStyle}

\newcommand{\PM}{\ensuremath{\mathcal{P}}\xspace}	
\newcommand{\PD}{\ensuremath{P}\xspace}  					

\newcommand{\dt}{\ensuremath{\dif \TimeIndex}\xspace}  
\newcommand{\dx}{\ensuremath{\dif x}\xspace}	
\newcommand{\dPM}{\ensuremath{\dif \PM}\xspace}	

\newcommand{\ND}[2]{\ensuremath{\mathcal{N}\left(#1,#2\right)}\xspace}		
\newcommand{\NN}{\ensuremath{\mathbb{N}}\xspace}		
\newcommand{\reals}{\ensuremath{\mathbb{R}}\xspace}		

\newcommand{\MCSize}{\ensuremath{M}\xspace}

\newcommand{\TimeIndex}{\ensuremath{t}\xspace}
\newcommand{\Time}{\ensuremath{T}\xspace}

\newcommand{\EnsembleIndex}{\ensuremath{n}\xspace}
\newcommand{\EnsembleSize}{\ensuremath{N}\xspace}

\renewcommand{\vec}[1]{%
	\ensuremath{\mathbold{#1}}							
}

\renewcommand{\matrix}[1]{\ensuremath{\mathbf{#1}}}			

\newcommand{\distance}{\ensuremath{\vec{d}}\xspace}		
\newcommand{\ObjF}{\ensuremath{J}\xspace}							
\newcommand{\Moment}{\ensuremath{m}\xspace}						
\newcommand{\MomentOrder}{\ensuremath{q}\xspace}
\newcommand{\MomentsVecSize}{\ensuremath{K}\xspace}			
\newcommand{\momVec}{\ensuremath{\vec{m}}\xspace}
\newcommand{\momVecEmp}{\ensuremath{\momVec^{\text{emp}}}\xspace}			
\newcommand{\momVecSim}{\ensuremath{\momVec^{\text{sim}}}\xspace}		
\newcommand{\WeightMat}{\ensuremath{\matrix{W}}\xspace}									
\newcommand{\paramVec}{\ensuremath{\vec{\theta}}\xspace}							
\newcommand{\paramVecTrue}{\ensuremath{\paramVec_0}\xspace}							
\newcommand{\paramVecALW}{\ensuremath{\paramVec^{\text{ALW}}}\xspace}	
\newcommand{\paramVecFW}{\ensuremath{\paramVec^{\text{FW}}}\xspace}		
\newcommand{\paramVecEst}{\ensuremath{\hat{\paramVec}}\xspace}				

\newcommand{\Tshort}{\ensuremath{\Time_{\text{short}}}\xspace}
\newcommand{\Tlong}{\ensuremath{\Time_{\text{long}}}\xspace}
\newcommand{\MixNT}{\ensuremath{\Time \times \EnsembleSize }\xspace}	

\newcommand{\DemandF}{\ensuremath{D_\text{f}}\xspace}				
\newcommand{\DemandFt}{\ensuremath{D_{\text{f},t}}\xspace}	
\newcommand{\DemandC}{\ensuremath{D_\text{c}}\xspace}				
\newcommand{\DemandCt}{\ensuremath{D_{\text{c},t}}\xspace}	
\newcommand{\Fundis}{\ensuremath{N_\text{f}}\xspace}				
\newcommand{\FundisT}{\ensuremath{N_{\text{f},t}}\xspace}		
\newcommand{\Chartists}{\ensuremath{N_\text{c}}\xspace}			
\newcommand{\ChartistsT}{\ensuremath{N_{\text{c},t}}\xspace}
\newcommand{\SigmaF}{\ensuremath{\sigma_\text{f}}\xspace}		
\newcommand{\SigmaC}{\ensuremath{\sigma_\text{c}}\xspace}		
\newcommand{\Vf}{\ensuremath{V_\text{f}}\xspace}						
\newcommand{\Vft}{\ensuremath{V_{\text{f},t}}\xspace}				
\newcommand{\Vc}{\ensuremath{V_\text{c}}\xspace}						
\newcommand{\Vct}{\ensuremath{V_{\text{c},t}}\xspace}				
\newcommand{\FracFundis}{\ensuremath{n_\text{f}}\xspace}					
\newcommand{\FracFundisT}{\ensuremath{n_{\text{f},t}}\xspace}			
\newcommand{\FracChart}{\ensuremath{n_\text{c}}\xspace}						
\newcommand{\FracChartT}{\ensuremath{n_{\text{c},t}}\xspace}			
\newcommand{\epsFundis}{\ensuremath{\epsilon_\text{f}}\xspace}		
\newcommand{\epsFundisT}{\ensuremath{\epsilon_{\text{f},t}}\xspace}
\newcommand{\epsChart}{\ensuremath{\epsilon_\text{c}}\xspace}			
\newcommand{\epsChartT}{\ensuremath{\epsilon_{\text{c},t}}\xspace}

\newcommand{\threeS}{\ensuremath{^{***}}\xspace}			
\newcommand{\twoS}{\ensuremath{^{**}}\xspace}					
\newcommand{\oneS}{\ensuremath{^{*}}\xspace}					
\title{
Time is limited on the road to asymptopia%
\thanks{For valuable feedback we thank participants of the Ergodicity Economics 2022 Conference and the virtual seminar on \textit{Quasi-Ergodic Measures} jointly organized by Zachary Adams (MPI MiS) and Maximilian Engel (FU Berlin).}
}

\subtitle{
An analysis of the ergodic properties of moment functions in the validation of financial agent-based models
}

\author[1]{\person{Ivonne Schwartz}\thanks{\texttt{\href{mailto:ivonne.schwartz@mailbox.org}{~\Letter~ivonne.schwartz@mailbox.org}}
Funded by the Hans-Böckler-Stiftung (PK045) in the Bamberg Doctoral Research Group on Behavioral Macroeconomics (BaGBeM).
}}
\author[2]{\person{Mark Kirstein}\thanks{\texttt{\href{mailto:mark.kirstein@mis.mpg.de}{~\Letter~mark.kirstein@mis.mpg.de}}}}
\affil[1]{University of Bamberg, Department of Economics, Germany}
\affil[2]{Max Planck Institute for Mathematics in the Sciences, Leipzig, Germany}

\begin{document}

\maketitle

\begin{abstract}
\noindent
One challenge in the estimation of financial market agent-based models (FABMs) is to infer reliable insights using numerical simulations validated by only a single observed time series.
Ergodicity (besides stationarity) is a strong precondition for any estimation, however it has not been systematically explored and is often simply presumed.
For finite-sample lengths and limited computational resources empirical estimation always takes place in pre-asymptopia.
Thus broken ergodicity must be considered the rule, but it remains largely unclear how to deal with the remaining uncertainty in non-ergodic observables.
Here we show how an understanding of the ergodic properties of moment functions can help to improve the estimation of (F)ABMs.
We run Monte Carlo experiments and study the convergence behaviour of moment functions of two prototype models.
We find infeasibly-long convergence times for most.
Choosing an efficient mix of ensemble size and simulated time length guided our estimation and might help in general.
\end{abstract}
\vspace{1em}
\vspace{.5em}

\noindent\textsf{\textbf{JEL Classification\ }}
\href{https://www.aeaweb.org/econlit/jelCodes.php?view=jel#C}{
C61}	
$\cdot$
\href{https://www.aeaweb.org/econlit/jelCodes.php?view=jel#D}{%
D01 	
$\cdot$
D81 	
$\cdot$
D9} 	
%


\newpage


\section{Introduction}

Financial crashes, global pandemics or disruptive innovations all qualify as formative events during which economic systems reach a new order.
On an aggregate level such events can trigger the transition from an old equilibrium to a new equilibrium.
A rapid sequence of such formative events can prevent the economy, financial markets and observables from ever reaching their long-time equilibria and thus keeping them steadily out-of-equilibrium.
In an effort to use a new methodology to understand such complex emergent behaviors and answer urging policy questions\autocite{Bouchaud2008,FarmerFoley2009,LuxWesterhoff2009} a community of researchers adopted financial agent-based models (FABMs) that deliberately allow for more complexity than traditional models.\autocite{TesfatsionJudd2006,HommesLeBaron2018}
The scientific challenge is to work with FABMs that are capable to endogenously reproduce known statistical regularities and that remain statistically and computationally tractable for estimation exercises.\footnote{The most important stylized facts of financial time series include phenomena like excess volatility, fat-tailed return distributions, absence of autocorrelations in the raw returns but long memory in absolute returns and volatility clustering. Regarding the study of financial markets, there exists a large strand of literature which provides models that explain the stylized facts well. Surveys about the statistical properties of financial markets can be found in \textcite{Cont2001,LuxAusloos2002,Hommes2006,LeBaron2006,Lux2009}.}
The ever increasing amounts of data and computational resources make FABMs a promising candidate to ultimately join the set of models which drive informed policy decisions.\autocite{WesterhoffFranke2018}

The development of a consensual validation and estimation protocol for (F)ABMs is important for scientific standards of reproducibility.
However, the estimation and validation of FABMs remains challenging and despite a growing body of research there exists no consensus in the literature, yet.\footnote{\textcite{RichiardiEtAl2006} is one attempt in this direction, \textcite{DelliGattiEtAl2018} is a helpful textbook. See \textcite{FagioloEtAl2019,LuxZwinkels2018} for recent surveys.}
Having a standardized protocol after all would address main methodological critiques raised against agent-based modeling.\autocite{LeombruniRichiardi2005,RichiardiEtAl2006,FagioloRoventini2017}

Here we follow a direction outlined in \textcite[96]{GaffeoEtAl2007} of a \textquote{descriptive output validation}, where we focus on simulated moment estimators which have been studied already intensively for FABMs\autocite{Amilon2008,GilliWinker2003,WinkerEtAl2007,Franke2009,Schmitt2021,Schwartz2022}.
Apart from the simulated method of moments (SMM) used here, different estimation approaches have recently emerged in the literature including likelihood-based methods, information-theoretic similarity-based measures and more recently a number of Bayesian estimations of FABMs appeared, too.\autocite{KukackaBarunik2017,Lamperti2018,Barde2016,GrazziniEtAl2017,Platt2021,BertschingerMozzhorin2021,DelliGattiGrazzini2020,Lux2018,Lux2021,Lux2021a}

Reliable estimation rests on sufficient convergence speed given finite-sample lengths and limited computational resources.
Surprisingly, the development of better tools and protocols in the validation of (F)ABM has often overlooked the ergodicity issue so far.
Our contribution enters exactly at this issue.
Already the seminal contribution by \textcite[225]{Cont2001} emphasised ergodicity in the process of estimation of parameters in models of financial market:
\blockquote[emphasis added]{One needs an ergodic property which ensures that the time average of a quantity converges to its expectation. \emph{Ergodicity} is typically satisfied by IID observations but it is \emph{not obvious -- in fact it may be very hard to prove or disprove} -- for processes with complicated dependence properties such as the ones observed in asset returns.}
Admittedly, ergodicity is an elusive property and the fact that no ready-to-use econometric test for ergodicity exists stands in the way of a comprehensive treatment.
It may come as less of a surprise then that only few publications identify ergodicity as an important assumption at all.\footnote{To convey a rough idea, the four volumes of the \citetitle{TesfatsionJudd2006} only casually mention 'ergodic'; esp. in the volumes most relevant, \ie Vol. 2 on \textit{\citefield{TesfatsionJudd2006}{subtitle}} \parencite{TesfatsionJudd2006} and Vol. 4 on \textit{\citefield{HommesLeBaron2018}{subtitle}} \parencite{HommesLeBaron2018}, 'ergodic' is mentioned in Ch. 22 \& 32 and resp. in Ch. 8 \& 14, however with no relation to the validation of ABMs.}
Regardless, ergodicity often lurks in the background as an implicit regularity condition.\autocites[19]{SoltykChan2021}[p. 25, footnote 36]{FrankeWesterhoff2016}[p. 91, footnote 8]{LeBaron2021}
As mentioned by \textcite{Cont2001} a proof of ergodicity of the relevant observables is not trivial, because it depends on the convergence speed and the respective time scales are often large.
The danger for modellers is to run into unnoticed trouble during the process of estimation and validation since knowledge about the estimator's convergence behaviour is instrumental.

In this paper, we aim to contribute to a growing body of research on the validation and estimation of FABMs.
In particular, we add useful steps towards a coherent study of the effects of broken ergodicity.
We do not propose any new estimation technique, but rather enhance the understanding and applicability of existing ones with regard to the ergodicity issue.
Our main contributions are the following.
First, there exists a nucleus of publications studying ergodicity in the estimation of (F)ABMs especially by \person{Jakob Grazzini} and \person{Matteo Richiardi}, which started with the study of models much simpler than the FABMs we study here\autocite{Grazzini2011,GrazziniRichiardi2013,GrazziniRichiardi2015}.
We analyse broken ergodicity in the SMM estimation of two well established FABMs, the \textcite{AlfaranoEtAl2008} (\ALWmodel) and the \textcite{FrankeWesterhoff2012} (\FWmodel), which reproduce a much larger set of stylized facts of financial markets.
We perform Monte Carlo simulations and use the simulated method of moments (SMM) approach for the estimation of both ABMs.\footnote{Standard references for the SMM are \textcite{McFadden1989,PakesPollard1989,LeeIngram1991,DuffieSingleton1993}. A textbook treatment is in \textcite[Chap. 9.6]{DavidsonMacKinnon2003}.}

Second, we use the terminology of time averages and ensemble averages which allows to express many challenges in the estimation process as depending on taking different types of averages.
For instance, we explicitly analyse the convergence behaviour over time vis-a-vis the convergence behaviour over replications.
Doing so, we show how to reduce the uncertainty coming from broken ergodicity by analysing the estimation for different choices of ensemble size and simulation time length under limited computational resources, for which most estimators remain in a pre-asymptotic regime.

A brief survey of related work puts our contribution in context of the existing literature.
Our work is closely related to a series of publications initiated by \person{Grazzini} and \person{Richiardi} who in \textcite{Grazzini2012,GrazziniRichiardi2015} introduced a useful terminology and also use the SMM approach.
However, they analyse the ergodic properties of much simpler models than those FABMs that we study here.
As stated by \textcite[154]{GrazziniRichiardi2015}, ergodicity assures consistent estimates even in the adjustment phase, \ie when the system has not yet fully reached the limiting distribution, because the properties are invariant across different realizations of the stochastic process.
This assumes that certain regularities are \textquote{stable enough across different replications of the model} to be identified.
In general, \textcite{Grazzini2012,GrazziniRichiardi2015} use non-parametric statistical tests to test for ergodicity (and stationarity). \textcite{Grazzini2012} applies a Wald-Wolfowitz runs test to check whether the observed and simulated data come from the same population.
This can also be done using a Kolmogorov-Smirnov test.\footnote{For ergodic Markov processes, the tests by \textcite{DomowitzElGamal1993,DomowitzElGamal2001} are applicable.}
These tests can offer a first starting point.
However, \textcite{DosiEtAl2018} find conflicting results using these econometric tests and thus conclude that they are not reliable to detect (non-)ergodicity.
Another approach based on robust statistics is given by \textcite{WinkerEtAl2007} who study situations when a typical distribution or time series is better represented by the median instead of the mean.
The divergence of the median from the mean is a typical symptom of broken ergodicity as the mean or ensemble average is prone to outliers and because of that tracks the \textit{exceptional} individual whereas the median is resistant to extremes and approximates the time-averaged behaviour and thus the behavior of \textit{typical} individuals.

Furthermore, sufficient asymptotic convergence might be beyond reach for all practical purposes.
\textcite{BouchaudFarmer2021} recently termed these situations \textit{quasi non-ergodicity}, \ie convergence times are astronomically-long such that for most practical purposes the estimation takes place in a pre-asymptotic regime where ergodicity is notoriously hard to achieve.
Similarly \textcite{DessertaineEtAl2022} draw another analogy from physics and refer to non-ergodic effects of estimation in pre-asymptopia as violations of the \textit{adiabatic assumption}, \textquote{in the sense that the time needed for the system to reach equilibrium is much shorter than the time over which the environment changes, so out-of-equilibrium effects can be neglected}.
The model data-generating process (mDGP) of a (F)ABM is usually not analytically solvable let alone available in closed-form, which makes it even harder to choose the number of simulation runs and length without any prior knowledge about the speed of convergence.
Our study helps to reduce this uncertainty.

The remainder of the paper is organised as follows.
In \secref{Method} we provide a basic understanding of ergodicity for the purpose of moment selection.
It turns out that the simulated method of moments like any estimation approach closely relies on ergodicity somewhere along the process.
\secref{Method} shows where exactly ergodicity enters.
In \secref{SimulationExperiments} we investigate how (broken) ergodicity influences the validation of FABMs.
In doing so, we run a bunch of numerical experiments on two financial market models and present the results of our simulations.
\secref{Conclusion} contains a summary and concludes the paper.

\section{Ergodicity and moment selection} \seclabel{Method}
This section starts with a brief discussion of ergodicity for the purpose to guide the selection of proper moments.
In short, proper moments must be ergodic to be invariants of the model and not artefacts of particular simulated trajectories.
With the help of an example of (geometric) Brownian motion we show that it is possible and necessary to derive ergodic observables from non-ergodic processes.
These insights about ergodicity then guide the process of moment selection in \ssref{MomentSelection}.
The moments enter the method of simulated moments which is briefly described in \ssref{SMM}.

\subsection{Ergodicity}

Ergodic theory can be understood as a mathematical theory that studies the asymptotic behaviour of averages.
Results from this mathematical field become relevant in an economics context whenever an economic observable is modelled as a stochastic process,
In our context of financial markets, the asset price as our economic observable of interest is modelled as a random variable in a static context or as a stochastic process if we are interested in the time evolution as well.
\Fref{ErgodicAveraging} provides an intuitive understanding of the core aspects of ergodicity and shows five sample time series. 

\begin{figure}[h]
  \begin{tikzpicture}[>=stealth']
  	\node at (0,0) {\includegraphics[width=\linewidth]{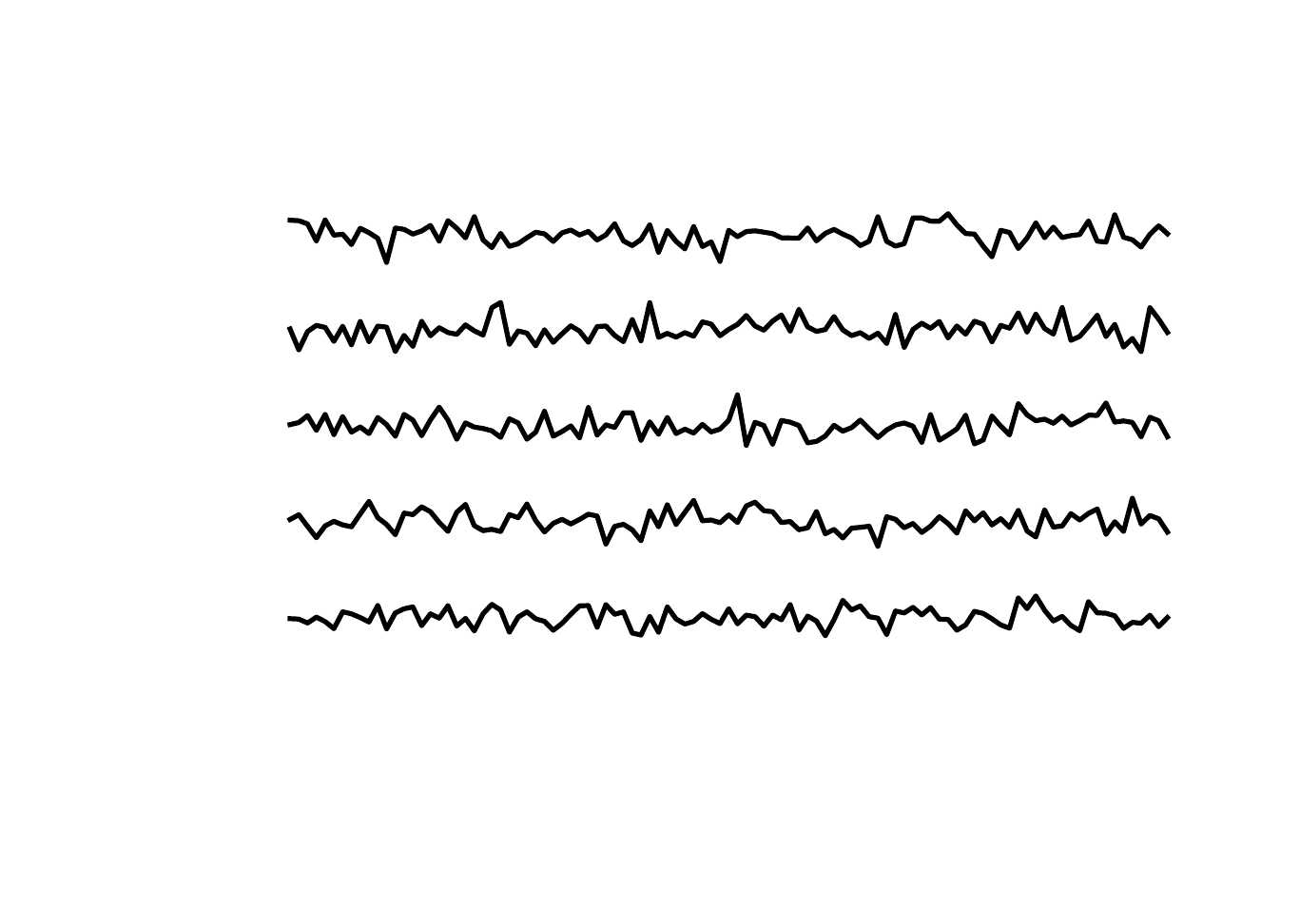}};
  	\draw[left,dotted] (-4.3,3) -- (-4.3,-3);
  	\node[below] at (-3,-3)	{start of observation};
  	\draw[dotted] (6.1,3) -- (6.1,-3);
  	\node[below] at (5,-3)				{end of observation};
  	\draw [->,ultra thick,color=red] (-5,3.7) -- (7,3.7);
   	\node[above] at (0.0,3.9)	{time $t$};
   	\draw [->,ultra thick,color=blue] (-6.5,3) -- (-6.5,-3);
   	\node[above,rotate=90] at (-6.7,0)	{realizations or universes \EnsembleIndex};
  	\node[right] at (-6.4,2.5)			{$X_t\left(x_1\right)$};
  	\node[left] at (-4.2,2.5)				{$\cdots$};
  	\node[right] at (6.1,2.5)				{$\cdots$};
  	\node[right] at (-6.4,1.4)			{$X_t\left(x_2\right)$};
  	\node[left] at (-4.2,1.4)				{$\cdots$};
   	\node[right] at (6.1,1.4)				{$\cdots$};
  	\node[right] at (-6.4,0.3)			{$X_t\left(x_3\right)$};
  	\node[left] at (-4.2,0.3)				{$\cdots$};
  	\node[right] at (-6.4,-0.8)			{$X_t\left(x_4\right)$};
  	\node[left] at (-4.2,-0.8)			{$\cdots$};
  	\node[right] at (6.1,-0.8)			{$\cdots$};
  	\node[right] at (-6.4,-1.9)			{$X_t\left(x_5\right)$};
  	\node[right] at (6.1,-1.9)			{$\cdots$};
		\node[left] at (-4.2,-1.9)			{$\cdots$};
  	\node at (-5.8,-2.6)						{$\vdots$};
  	\filldraw[nearly transparent,color=red,rounded corners] (-4.2,0.8) rectangle (5.9,0.1);
  	\draw[->,ultra thick, color=red] (5.9,0.4) -- (6.5,0.4);
  	\node[right] at (6.5,0.3)			{$\TA{X}_{x_3,\bullet}$};
  	\node[below] at (5.2,-3.8)			{$\TA{X}_{x_3,\bullet} = \lim\limits_{\Time\to\infty} \frac{1}{\Time} \sum\limits_{t=0}^{\Time-1} X_{x_3,t}$};
  	\filldraw[nearly transparent,color=blue,rounded corners] (0,-3) rectangle (0.5,3.2);
 		\draw[->,ultra thick,color=blue] (0.25,-3) -- (0.25,-3.8);
  	\node[below] at (0.25,-3.8)		{$\EA{X_{\bullet,t^*}} = \lim\limits_{\EnsembleSize\to\infty} \frac{1}{\EnsembleSize} \sum\limits_{n=1}^{\EnsembleSize} X_{x_n,\TimeIndex^*}$};
  \end{tikzpicture}
\caption{\textbf{Two Averaging Procedures.} Averaging over all realizations at a fixed time, \eg at $t^*$, yields the ensemble average at that time (blue shade): $\EA{X_{\bullet,t^*}} = \lim_{\EnsembleSize\to\infty} \nicefrac{1}{\EnsembleSize} \sum_{n=1}^{\EnsembleSize} X_{x_n,\TimeIndex^*}$.
	Averaging over time of a single realization of the stochastic processes, \eg the sample path associated with initial condition $x_3$, yields the time average (red shade): $\TA{X}_{x_3,\bullet} = \lim_{\Time\to\infty} \nicefrac{1}{\Time} \sum_{\TimeIndex=0}^{\Time-1} X_{x_3,\TimeIndex}$ \parencite[adapted from][68]{Kirstein2019}.}
\flabel{ErgodicAveraging}
\end{figure}

From a modelling standpoint we think of the five time series as five realizations of a stochastic process $\left\{ X_t \right\}_{t\in\TT}$ with five different initial conditions $x_1, \ldots, x_5$.%
\footnote{For simplicity we will omit the initial conditions in the notation if they are not relevant.}
For FABMs, the raw observables will be time series of asset prices or returns initialized by the series' first observations.
We think of the observed data as being generated by some unknown real-world data-generating stochastic process (rwDGP), which remains a purely mental construction, of course.
The FABM is the model data-generating process (mDGP), from which we are able to sample.
Validation and estimation in econometrics can then be understood as a test of the match between the mDGP and the rwDGP with respect to some metric.
In SMM, this metric consists of a vector of moments.
The selection of moments is discussed in \ssref{MomentSelection}.

With the help of ergodicity and \fref{ErgodicAveraging}, we distinguish between two types of the physical identity of an average, ensemble average (blue shade) and time average (red shade).
Ensemble and time averages perform different aggregations on different instances of the observables.
A time average computes an average along a single time series (or a single realization of a stochastic process) which is is depicted in \fref{ErgodicAveraging} by the red shade tracking the time series $X_t(x_3)$.
An ensemble average computes an average over different realizations of the random variable (or stochastic process) at a single point in time which is depicted in \fref{ErgodicAveraging} by the blue shade covering realizations at time $t^*$.
In econometric terminology time averages correspond to longitudinal averages and ensemble averages to cross-sectional averages.
Intuitively, a time average gives an aggregate statement about a single system over time, an ensemble average gives an aggregate statement about many systems at an instance in time.%
\footnote{A key insight that has recently become better understood is to distinguish not only between the physical identity of an average (ensemble/time) but to further abstract and distinguish the physical identity from the mathematical identity or mathematical functional form of an average (arithmetic/geometric/harmonic/\etc) \parencite{Kirstein2019}.}

Ergodicity is then a mathematical property of an observable that describes the conditions under which both averages exist and coincide.
A common definition of ergodicity establishes some type of convergence of longitudinal or time series averages on the LHS towards the cross-sectional average on the RHS:

\begin{equation} \elabel{ErgodicTheorem}
	\underbrace{\lim_{\Time\to\infty} \frac1\Time \sum_{\TimeIndex=1}^\Time f\left(x_\TimeIndex\right)}_{\text{average along a single infinite time series}}
	\overset{\text{a. s.}\strut}{\longrightarrow}
	\underbrace{\int_\Omega f(x)P(x) \dif x
	\vphantom{\sum_{\TimeIndex=1}^\Time}		
	}_{\text{average over population}} = \EA{X}	~,
\end{equation}
where $f$ is an arbitrary function of the observables.
This handy notation allows to cover the case of analysing the raw observables as well as some transformation of them, in the former case $f$ is simply the identity mapping, $f(x) = x$.
In our context, the function $f$ potentially represents some summary statistic of the data, \ie a moment function which we discuss below in more detail.
The LHS of \eref{ErgodicTheorem} yields a time average in the long-time limit.
It is important to note that the RHS is also the result of a limit, it yields the ensemble average in the large-ensemble limit.
The latter is rarely made explicit and instead the RHS is often presented just as \textit{the} expectation value, \ie \EA{X} = \EV{X}.

\subsection{Identifying ergodic observables}
In this section we briefly discuss the possibility to extract ergodic observables from non-ergodic processes.
This corresponds to extracting moment functions from some rwDGP of which we don't need to know whether it is ergodic or not as long as we select an ergodic moment function.
The goal is to identify invariants of a model which is a key goal in science.\autocite{Petersen1996}
Invariants are adequate quantities when describing and comparing models of real-world phenomena.
Identifying these invariants and package them in moment functions will be instrumental for the selection of proper moments discussed in the subsequent section.

Ergodicity as a property of mathematical models characterises such possible invariants in the specific sense of \eref{ErgodicTheorem}, \ie an ergodic observable converges (almost always) towards a unique invariant limit distribution.\footnote{In a sense, an ergodic observable ensures that more data is better for estimation and validation purposes, because more observations are likely to lead to better convergence.}
Most models of real-world phenomena are \perse non-ergodic processes, because some dynamic aspect (like growth) is relevant.\footnote{Let us emphasise that ergodicity is not a property of reality but of the mathematical model.}
For illustration purposes let us look at the standard model of asset price evolution which is that of (geometric) Brownian motion (GBM).

As such the model of GBM is clearly non-ergodic.
Yet it is possible to derive ergodic observables from this non-ergodic process.
Quite generally, a stochastic process is a model of an uncertain economic observable that can vary in time and over realizations.
A simple stochastic process is the Wiener process $W_t$.
In finance the Wiener process is used as a model of the price evolution, \ie the repeated measurements of the price of an asset.%
\footnote{The Wiener process is sometimes also called \textit{Brownian motion} because it was initially conceived to model the real-world phenomenon of Brownian motion, \ie the jittery motion of pollen suspended in fluids.}
The Wiener process $W_t$ is non-ergodic, as its time average behaves as $\TA{W_t} \sim \ND{0}{\nicefrac{t}{3}}$ for $t\to\infty$, and thus does not converge, because the distribution has a time-dependent variance, \ie the distribution is broadening over time.
Therefore, the \textit{time-dependent} time average cannot almost always be equal to its \textit{time-independent} ensemble average $\EA{W_t} = 0$, if so then only occasionally (by chance).

Again, we are interested in identifying ergodic observables or summary statistics of a process.
In the unlikely case that the raw observable is ergodic we are lucky.
More generally some transformations might be necessary to arrive at an ergodic observable.
Typical derived observables are increments, squared increments (sometimes referred to as squared displacements), correlations, multiples, returns or growth rates.
If we analyse the squared displacement from the origin of a Wiener process as our observable of interest, $D_T^2 = \left(W_T - W_0 \right)^2$, then this observable is ergodic, $\EA{D_T^2} = \TA{D_T^2}$.\autocite{MetzlerEtAl2014a}
The increments of a Brownian motion converge towards a unique invariant time-independent distribution, and are in this sense an ergodic observable.
The increments of a geometric Brownian motion, however, are not converging but remain level-dependent.
For GBM it is the exponential growth rate that converges to an ergodic invariant distribution instead.
This little example is meant to show that it is possible to derive ergodic observables from non-ergodic processes.
In fact, we use the ergodic observables to characterize non-ergodic processes and thus in a similar way like in this little example use ergodic moments to characterize the mDGP of a (F)ABM.

It is the ergodic property that makes a summary statistic a suitable moment for the validation of (F)ABMs.
However, the high degrees of freedom in (F)ABMs often prevent the existence of a closed-form solution, which would simplify the direct analysis of the stochastic data-generating process and to analytically derive the limit distributions of observables.
Therefore, we have to rely on numerical simulations to study the convergence behavior and ergodic properties of moment functions.
With this in mind, we now turn to the process of moment selection.

\subsection{Selection of moment functions}\sseclabel{MomentSelection}
The term \textit{moment} is central in all moment-based estimations as the generalized method of moments (GMM) and the simulated method of moments (SMM) which we use in this paper.
However, the word \textit{moment} has different meanings in different fields.
Let us therefore briefly clarify how we use it here.
Whenever we refer to \textit{moments}, we use this term in the narrow sense of their ordinary meaning in stochastics as properties of a probability distribution \PM in \eref{MomentsMeasure}.
In the context of SMM estimation of (F)ABMs, the literature refers to moments in a wider sense as some summary statistic.
To avoid confusion we will thus refer to them as \textit{moment functions} in this paper.

\subsubsection{Moments}
In stochastics a moment \Moment of order $\MomentOrder = 0,1,2,3, \ldots ~; \MomentOrder\in\NN$ of a distribution is computed as the mean of the \MomentOrder-th power of its realizations, $\Moment_\MomentOrder = \int_{-\infty}^{\infty} x^\MomentOrder \PD(x) \dx = \EV{X^\MomentOrder}.$
It is often convenient and therefore desirable to work with probability density functions such as \PD, but the density of a measure doesn't always exist, thus we can express moments using only the probability measure \PM instead,
\begin{equation}	\elabel{MomentsMeasure}
	\Moment_\MomentOrder = \EV{X^\MomentOrder} = \int_{-\infty}^{\infty} x^\MomentOrder \dPM(x) ~ .
\end{equation}

As the mDGP for most FABMs is not readily available in closed form, the moment functions can't simply be calculated analytically.
Thus we need to simulate the data first and then compute the moments numerically.
Inference from a given sequence of moments to a distribution is a classical problem in mathematics known as the moment problem. 
The estimation task of a (F)ABM is an inverse form of the classical moment problem.
Generally, moment problems are only solvable if long sequences of moments are available, which is usually not the case for distributions of observables of financial markets because they turn out to be heavy-tailed.%
\footnote{A classic text on the moment problem is \textcite{Akhiezer1965}, more recent research reveals the solvability of the \person{Hamburger} moment problem depends on the non-negative extendability of the \person{Hankel} matrix, see \textcite{Bolotnikov1996,Bolotnikov2008,ChenHu1998,DyukarevEtAl2008}.}
Higher-order moments of heavy-tailed distributions (HTDs) quickly vanish and in extreme cases even the first moment doesn't exist.\autocite{Rachev2003}
It seems therefore unlikely that solutions to classical moment problems are of immediate help for our estimation task.
However, it provides an additional motivation to extend the selection of moment functions in SMM beyond the ordinary moments.%
\footnote{\textcite{SoltykChan2021} use solutions to classical moment problems for modelling time-varying higher-order conditional moments of financial time series.}

Empirical analyses of economic and financial data at least since the 1960s\footnote{\textcite{Mandelbrot1960,Mandelbrot1963,Fama1963} are prominent references that find heavy-tailed price increments (returns).} have repeatedly confirmed that many observables are not very well described by the family of \person{Gaussian} distributions but instead follow HTDs, often with a power law tail.
Since the increasing magnitude of an extreme event raised to the \MomentOrder-th power, $x^\MomentOrder$, can't be offset by its decreasing probability, $\dPM(x)$, the integral in \eref{MomentsMeasure} diverges  and higher-order moments cease to exist for HTDs.
For instance, the calculation of the second central moment, the variance, is numerically always possible but does not converge for larger samples.
Associated with the (existence of the) variance is a typical scale of dispersion.
If the distribution has no second moment, then such the variations are so extreme that they lack a typical scale.
Obviously, non-existing moments would not be good candidates for moment functions.
An invariant of such volatile behavior might be the scaling exponent of the increments or returns.
In fact, it is now understood that power laws are invariants of scaling relationships which again are imprints of complex systems and their otherwise out-of-equilibrium dynamics.

\subsubsection{Moment functions} \sseclabel{MomentFunctions}
Moments \textquote[{\cite[491]{Greene2018}}]{provide a natural source of information about the parameters, other functions of the data may also be useful}.
A natural motivation to move beyond moments to moment functions is to not only use static properties of distributions but also dynamic properties of the underlying processes, \eg correlations are not a property of a distribution but of observations at different times, locations or both.
We refer to these 'other functions of the data' as \textit{moment functions}.
Our notation proves convenient here, as it is possible to interpret the function $f$ in \eref{ErgodicTheorem} simply as some moment function $f(x) = m(x)$.
Eventually, our vector of moments \momVec can contain ordinary moments in the sense of \eref{MomentsMeasure} as well as moment functions.

\paragraph{A hierarchy of convergences in ergodic theorems}
In order to emphasise how strong a statement about an ergodic property actually is, it is worthwhile to make explicit a hierarchy of convergences contained in ergodic theorems like \eref{ErgodicTheorem}.
For an ergodic theorem to hold:
\begin{enumerate}
  \item almost all individual time series averages need to converge at all,
  \item almost all individual time series averages need to converge to some (arbitrary) identical value,
  \item and this value must be a particular one, namely coincide with the expectation value.
\end{enumerate}
Let us comment on this hierarchy.
On the first two points, most raw time series do not converge at all to a single value and are thus not stationary in this sense, \eg all growth processes, security prices, business cycle dynamics or exchange rates.
Some transformation like first or higher differencing is necessary to impose stationarity and thus convergence.\autocite[145]{DelliGattiEtAl2018}
However, such transformations do not necessarily impose ergodicity, as they operate only on a single time series.
On the third point, the expectation value is a particular average, \ie the probability-weighted arithmetic mean of all realizations of a random variable.
Thus its mathematical identity is an arithmetic mean and its physical identity is an ensemble average.
Note that in general both time and ensemble averages do not coincide and hence non-ergodicity or broken ergodicity is our default condition.
In fact, ergodicity only holds for very special cases or carefully chosen transformations of the observables.

This convergence hierarchy also applies to moment functions.
Already \textcites[805]{Franke2009}[918]{RugeMurcia2012} note that many empirical moment functions are computed as time averages of some function of the (time series) data,
$\nicefrac1T \sum_{t=1}^{\Time} \Moment \left( x_t \right).$
We only take the obvious next step to demand the ergodicity of moment functions in the selection process such that they converge towards a unique value or distribution and capture invariants of the mDGP.
So far it is only tacitly assumed in the literature that moment functions are ergodic but rarely explicitly analyzed, which implies that the calculation for (almost) all time series samples converges (almost) always towards a particular and identical value (namely the expectation value).
For FABMs that offer no closed-form analytical solution, the convergence of moment functions can only be assessed numerically which will be done in \ssref{PreAsymptoticsMomentFunctions}.
Thus, the hierarchy of convergences is directly informative in the context of the validation of ABMs.
Firstly, the time averages of the moment functions need to converge at all and additionally to some value, $\nicefrac1T \sum_{t=1}^{\Time} \Moment \left( x_t \right) \to \TA{\Moment}$.
Corresponding to the third aspect in the hierarchy, this convergence needs to go towards the expected value we observed empirically,
$\TA{\Moment} = \nicefrac1T \sum_{t=1}^{\Time} \Moment \left( x_t \right) \to \EA{\Moment}$.
Eventually, proper moment functions have to rely be ergodic invariants of the mDGP. 


Ergodic theory implies a further highly relevant aspect to the validation of FABMs which is its asymptotic nature.
Roughly speaking, ergodic theorems live in the land of asymptopia, while reality and the estimation of (F)ABMs take place in pre-asymptopia with finite samples.
For all practical purposes real-world data will always be limited even if the available computational resources are constantly expanding.
If there are only finite observations in the time series of the FABMs in pre-asymptopia then the time scales of the convergences will become crucially important sooner or later.
Any uncertainty about the LHS in \eref{ErgodicTheorem} vanishes only in the large-time limit.\footnote{
But there is also some uncertainty about the true expectation value on the RHS in \eref{ErgodicTheorem}, which only vanishes in the large-ensemble limit.}
In pre-asymptopia all quantities are computed under the constraint of finite computational resources.
Thus we operate with samples finite in time and ensemble size, $\lim_{t\to\Time}, \Time \ll \infty$ and $\lim_{n\to\EnsembleSize}, \EnsembleSize \ll \infty$. 
Given any limited budget of computational resources, the uncertainties in computing moment functions of our FABM depend on such limits and behave differently and might never vanish sufficiently fast.

\subsection{Monte Carlo cube} \sseclabel{MCCube}
So far we have discussed (i) ergodicity in general, (ii) the importance of the ergodic property, (iii) assumptions hidden in the ergodic theorem about specific convergences that reappear in the selection of suitable moments of FABMs and (iv) the fact that validation of FABMs always takes place in pre-asymptopia where ergodicity is generically broken due to slow convergences.
In this section we focus on the different dimensions where convergences take place when performing Monte Carlo simulations and how to visualize them.
This section prepares for a better understanding of the effects of broken ergodicity along the different dimensions as discussed in \sref{SimulationExperiments}.
It will turn out to be instructive for the choice of how to split a limited number of total observations over (i) the simulation length, (ii) the number of simulation runs and (iii) the number of Monte Carlo repetitions  to improve validation efforts.

We model observables as realizations of a random variable $X$.\footnote{Throughout this paper we denote random variables by capital letters and their realizations by the respective lower case.}
Repeated observations or realizations $x$ of the same random variable at regular intervals from time 0 to time $T$ form a time series of our observable, $\left\{ x_0, x_1, x_2, \ldots , x_T \right\}$, also denoted by $\left\{x_t\right\}_0^T$.
Within this stochastic model a specific observed time series can therefore be interpreted as a realization of finite length of a stochastic process $\left\{ x_t \right\}_{t\in\TT}$, where \TT denotes the time domain of the time index.%
\footnote{The flexible notation of the time domain as \TT allows to easily adopt different time domains.
For a continuous time model the time parameter $t$ of the stochastic process is $\TT = \reals$ or $\left\{ x_t \right\}_{t\in\reals}$.
For a model in discrete time $\TT = \NN$ which yields the stochastic process $\left\{ x_t \right\}_{t\in\NN}$.}
A stochastic process, $\left\{ x_t \right\}_{t\in\TT}$, is then a family of all infinite time series.

Our notation establishes the following relation between the real world (which is observable) and our model.
Nature or the financial markets are thought of as a rwDGP which is unknown and of which we often observe only exactly one unique time series.
On the other hand we have an economic model of a financial market -- a FABM in our case -- which is our model data-generating stochastic process (mDGP), from which we can generate an ensemble of (at least in principle) arbitrarily many $N$ time series of (finite) length $T$.
At least two facts distinguish two time series generated by the same stochastic process.
First, the two time series (or realizations of the stochastic process) differ in their random initial conditions or random seeds.\autocite[39-40]{DelliGattiEtAl2018}
The dependence on some initial conditions is indicated in \fref{ErgodicAveraging} by the argument in the function, \eg $X_t\left(x_1\right)$ signifies a process $X_t$ that started in $x_1$.
Second, randomness in the noise realizations -- there are simply different realizations of the random variable that appear over time or are generated and used in the simulation.

Throughout the paper, the distinction between the time dimension and the ensemble dimension will be crucial.
We denote time by $T$ and denote the size of the ensemble by $N$.
A Monte Carlo simulation experiment is then fully determined by the 
\begin{enumerate}
  \item simulated time length \Time, which yields a $(1 \times \Time)$-matrix or row vector of size \Time;
  \item an ensemble of \EnsembleSize different generated time series, which yields a $(\EnsembleSize \times \Time)$-matrix;
  \item number of \MCSize Monte Carlo realizations, which yields \MCSize different $(\EnsembleSize \times \Time)$-matrices.
\end{enumerate}

\begin{figure}[htbp]
\begin{center}
\begin{tikzpicture}[
		every node/.style={
			anchor=north east, 			
			fill=white,							
			minimum width=1.4cm,		
			minimum height=7mm			
			}
	]
	\matrix (mA) [draw,matrix of math nodes]
	{
	    x^{(\MCSize)}(1,1)	& x^{(\MCSize)}(1,2)	& \cdots & x^{(\MCSize)}(1,\Time)\\
	    x^{(\MCSize)}(2,1)	& x^{(\MCSize)}(2,2)	& \cdots & x^{(\MCSize)}(2,\Time)\\
	    \vdots        & \vdots				& \ddots & \vdots      \\
	    x^{(\MCSize)}(\EnsembleSize,1)	& x^{(\MCSize)}(\EnsembleSize,2)	& \cdots & x^{(\MCSize)}(\EnsembleSize,\Time)\\
	};
	\matrix (mB) [draw,matrix of math nodes] at ($(mA.south west)+(5,1)$)
	{
	    x^{(2)}(1,1)	& x^{(2)}(1,2)	& \cdots & x^{(2)}(1,\Time)\\
	    x^{(2)}(2,1)	& x^{(2)}(2,2)	& \cdots & x^{(2)}(2,\Time)\\
	    \vdots        & \vdots				& \ddots & \vdots      \\
	    x^{(2)}(\EnsembleSize,1)	& x^{(2)}(\EnsembleSize,2)	& \cdots & x^{(2)}(\EnsembleSize,\Time)\\
	};
	\matrix (mC) [draw,matrix of math nodes] at ($(mB.south west)+(5,1)$)
	{
	    x^{(1)}(1,1)	& x^{(1)}(1,2)	& \cdots & x^{(1)}(1,\Time)\\
	    x^{(1)}(2,1)	& x^{(1)}(2,2)	& \cdots & x^{(1)}(2,\Time)\\
	    \vdots        & \vdots				& \ddots & \vdots      \\
	    x^{(1)}(\EnsembleSize,1)	& x^{(1)}(\EnsembleSize,2)	& \cdots & x^{(1)}(\EnsembleSize,\Time)\\
	};
	\draw[dashed](mA.north west)-- node[sloped,above] {Monte Carlo runs $\longrightarrow$} (mC.north west);
	\node [above,rotate=90] at (mC.west) {realizations $\rightarrow$};
	\draw[dashed](mA.south east)--(mC.south east);
 	\node [below] at ($(mC.south east)-(3,0)$) {time $\longrightarrow$};
 	\draw[color=red,rounded corners,thick] (-12.5,-7.6) rectangle (-5.3,-7);
 	\node[right] at (-4.9,-7)	{\red{time averaging}};
 	\draw[color=blue,rounded corners,thick] (-7.4,-4.6) rectangle (-5.4,-8);
 	\node[right] at (-4.9,-8) {\blue{ensemble averaging}};
	\node[right] at (-4.9,-5.7) {\DarkGreen{MC averaging}};
\coordinate (TopLeftCorner) at (-2.1,-0.2);
\coordinate (TopRightCorner) at (0.3,-0.2);
\coordinate (BottomLeftCorner) at (-7.5,-5.3);
\coordinate (BottomRightCorner) at (-5.1,-5.3);
	\draw[color=green!40!black,thick] (TopLeftCorner)--(TopRightCorner);
	\draw[color=green!40!black,thick] (TopLeftCorner)--(BottomLeftCorner);
	\draw[color=green!40!black,thick] (TopRightCorner)--(BottomRightCorner);
	\draw[color=green!40!black,thick] (BottomLeftCorner)--(BottomRightCorner);
\end{tikzpicture}
\end{center}
\caption{\textbf{Monte Carlo Cube.}
Every column depicts an ensemble, \ie a collection of $N$ realizations of the random variable $x$ at some instance of time.
Every row of a matrix depicts a time series, \ie realizations of the random variable $x$ in a particular realization given by the first argument over time $t$ which is denoted by the second argument.
Every matrix of such rows and columns depicts the outcome of a single Monte Carlo run denoted by the superscript number in parenthesis. 
Three possibly different dimension along which averaging and thus convergence can take place, the ensemble dimension (blue), the time dimension (red) and a third dimension along different Monte Carlo runs (green).}
\flabel{MonteCarloCube}
\end{figure}
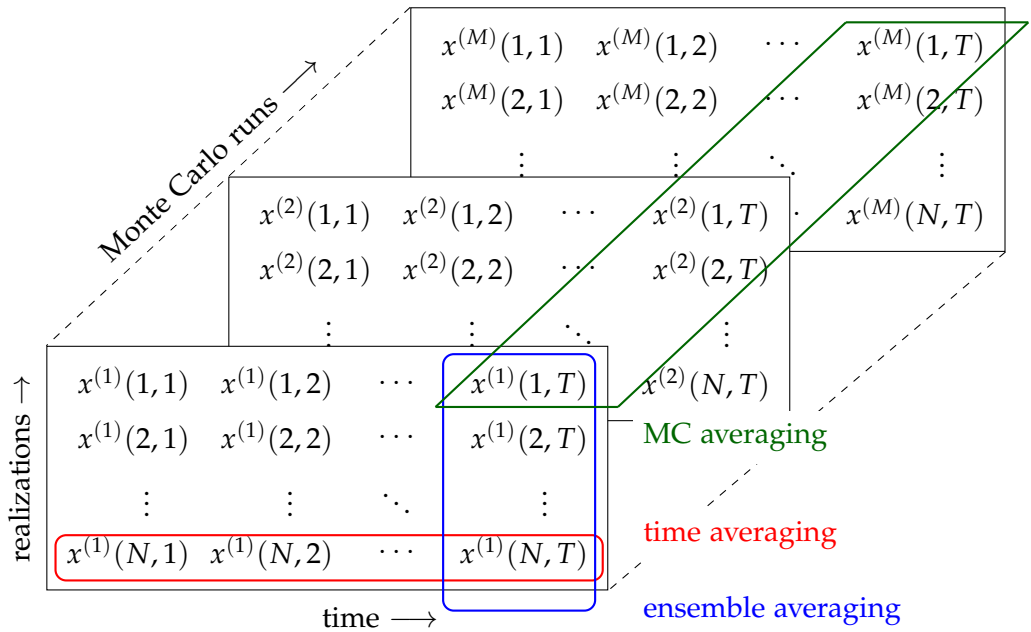

\Fref{MonteCarloCube} contains a visualization of the three dimensions of simulations of (F)ABMs.
Commonly, this third dimension of the Monte Carlo runs is understood as belonging to the ensemble dimension as only the random seeds might be different, but additional numerical effects appear that make a distinction between the two dimensions possible which are discussed in more detail in \sref{SimulationExperiments}.\footnote{Related two-dimensional visualisations in the context of tests for ergodicity can be found in \textcite[Fig. 3, p. 132]{GueriniMoneta2017}.
\textcite[155]{GrazziniRichiardi2015} refer to the three-dimensional mental model of a cube and refer to it as the \textquote{replications} dimension.
Similar reasoning about correct alignment of research question and statistical identification led to the idiographic paradigm in behavioral psychology and physiology and similar visualisations \parencite{MolenaarCampbell2009,NeumannEtAl2022}.}
The convergence behavior in the numerical estimation crucially depends on the ensemble size \EnsembleSize, the time length \Time and the number of Monte Carlo simulations \MCSize in a non-intuitive and non-linear way.
In principle different convergence behaviours in these three dimensions have not been investigated before.
Our analysis is thus contributing to the joint community efforts of improving the validation of (F)ABMs.


\subsection{Simulated method of moments} \sseclabel{SMM}

%
%

Let us now turn to the simulated method of moment estimation.
This estimation method is studied for different kind of models such as dynamic stochastic general equilibrium (DSGE) models\autocite{RugeMurcia2012,RugeMurcia2013} as well as (F)ABMs\autocite{Franke2009,FagioloRoventini2017,FagioloEtAl2019}.
In general SMM is more robust to misspecification than for example maximum likelihood methods\autocite{RugeMurcia2007} and performs better for large-scale models\autocite{Platt2020}.
SMM belongs to the broader class of simulated minimum distance methods whose goal is to identify the unknown parameters of the model, which generate the least distance between the simulated moment functions and the empirical observed moment functions.

Let us now introduce a simulated method of moment estimator.
Therefore, we consider a FABM with unknown parameter vector $\paramVec \in \Theta$, which we ultimately want to estimate from the parameter space $\Theta$.
In many empirical studies it is the primary goal to fit parameter values of the mDGP \eg by matching empirical and simulated data moment functions.
In our context, the goal is to assess the quality of the estimation approach itself, \ie how well does the estimation approach identify some known 'true' parameter values \paramVecTrue.
If no empirical data is used for the evaluation of the estimation approach but data simulated from an ABM -- like in our study -- then it is a common practice to use known parameter values from the literature 
as the benchmark of the estimation that match the moment functions well and refer to them as 'true' values.\footnote{They are sometimes also referred to as 'pseudo-true' values as they are derived from synthetic/simulated data. See also \sssref{TrueMomentVector}.}
Since SMM belongs to the broader class of simulated minimum distance methods, a distance function $\distance$ measures the difference between empirical and simulated data given some parameters \paramVec.\footnote{To be very precise, the empirical data is produced given the true parameters \paramVecTrue, which we want the estimation to recover reliably.}
Thus, we compute the distance between the two moment function vectors for the simulated and empirical moment functions $\distance = \momVecEmp\left(x_t\right) - \momVecSim\left(x_t|\paramVec\right)$ given some parameters \paramVec.
In the following, we explicitly consider two different types of averaging over simulated observables. First, simulated moment functions can be computed as time averages over one realization of observables $x_t$ of time length $\Time$:
\begin{equation} \elabel{momentConditionTA1}
	\momVecSim = \frac{1}{\Time} \sum_{t=1}^\Time m\left(x_t\right) ~.
\end{equation}
Or second, they can also be captured as averages over an ensemble of size \EnsembleSize of time averages of length \Time:
\begin{equation} \elabel{momentConditionTA2}
	\momVecSim = \frac{1}{\EnsembleSize\Time} \sum_{n=1}^\EnsembleSize \sum_{t=1}^\Time m \left(x_t\right) ~.
\end{equation}

The objective function or criterion function for a given a set of model parameters \paramVec aggregates then the distances
\begin{equation} \elabel{ObjectiveFunction}
	\ObjF \left( \paramVec \right) = \distance\left(\paramVec\right)^\prime \WeightMat~\distance\left(\paramVec\right) ~,
\end{equation}
for a given weighting matrix \WeightMat which is positive semi-definite.
The estimator \paramVecEst yields that vector of model parameters for which the objective function is minimized:
\begin{equation}
	\paramVecEst = \arg\min_{\paramVec} \ObjF \left(\paramVec\right) ~.
\end{equation}
Under standard regularity conditions, the distance function \distance is assumed to be stationary and ergodic resulting in an asymptotically consistent estimator.\autocite{LeeIngram1991,DuffieSingleton1993}
For most (F)ABMs, however, a corresponding SMM estimator might not have such properties.
In fact, we will show that in pre-asymptopia the following inequality holds due to non-commutativity of the limits in ensemble size, time and also of the number of MC runs:
\begin{equation}\elabel{ergodicInequality}
	\frac{1}{\EnsembleSize\Time} \sum_{n=1}^{\EnsembleSize} \sum_{t=1}^{\Time} \Moment \left(x_t\right)
	\neq 
	\frac{1}{\EnsembleSize\Time} \sum_{t=1}^{\Time} \sum_{n=1}^{\EnsembleSize} \Moment \left(x_t\right)
	~.
\end{equation}
Put simply, under broken ergodicity and/or in pre-asymptopia the order matters in which the limits are taken.\autocite[eq. 2 on p. 6]{Grazzini2011}
We thus study the effect of an efficient allocation of a limited budget of observations and how to get an estimator with improved properties.

The efficiency of the SMM estimator is affected by the design of the weighting matrix \WeightMat.
The optimal choice is given by a weighting matrix with the smallest asymptotic covariance for the estimator.
One popular choice would be the use of the Newey-West estimator.
In our numerical study, we are interested in the uncertainty that can be solely associated with broken ergodicity.
Therefore, we will consider the inverse of the long-run covariance matrix of the true data as the optimal weighting matrix.


To summarize this section, we have discussed how the ergodic property plays a crucial role in moment selection.
As we will see next, convergences along the three different dimensions might behave differently due to broken ergodicity in pre-asymptopia.
How exactly shows the evaluation of the simulation experiments in the next section.

\FloatBarrier

\section{Simulation experiments}\seclabel{SimulationExperiments}

To conduct our study of (broken) ergodicity in the convergence of moment functions, we take two established financial agent-based models: \textcite{AlfaranoEtAl2008} (\ALWmodel) and \textcite{FrankeWesterhoff2012} (\FWmodel).
They are both based on a herding mechanism that has its roots in \textcite{Kirman1993}.
Both models are commonly used for estimation exercises in the community.\autocite{ChenLux2018,Lux2018,KukackaKristoufek2020,BertschingerMozzhorin2021}
They replicate most of the stylized facts mentioned above and require comparatively little computational resources despite their prototype-nature.

\subsection{Agent-based financial market models} \sseclabel{FABMs}
In this subsection we briefly discuss the key mechanisms underlying the \ALWmodel and the \FWmodel which are the basis of our numerical experiments hereafter.

\subsubsection{ALW model} \sseclabel{ALWmodelDescription}
The first model by \textcite{AlfaranoEtAl2008} incorporates a behavioural herding mechanism based on \textcite{Kirman1993} with precedent analysis in \textcite{AlfaranoEtAl2005}.
\footnote{For a simple variant of the herding model by \textcite{Kirman1993} with local interaction between the agents (or equivalently extensive transition rates) a closed-form exists for which analytical solutions of the time-variation of moments and some related quantities of interest exist can be computed (see \cite{AlfaranoEtAl2005} and esp. \cite[Section 4]{AlfaranoEtAl2008}).}


The ALW model assumes two types of financial speculators, fundamentalist and chartist traders. 
Fundamentalists' excess demand is given by $\DemandF = \Fundis \Vf \left(p^{\star}_{t}-p_t\right)$ where \Vf is the average demand of \Fundis fundamental speculators.
The fundamental value $p^{\star}_{t}$ is assumed to follow a random walk $p^{\star}_{t}=p^{\star}_{t-1}+\SigmaF \cdot \epsilon_{f,t}$ with $\epsilon_{f}\sim \ND{0}{1}$.
The excess demand of chartist traders is  given by $\DemandC = \Chartists \Vc x_t$. Chartists are in one of two opinion states, either optimistic or pessimistic.
A sentiment index $x_t$ is defined as $x_t = \nicefrac{2n_t}{\Chartists}-1$ with $n_t$ optimistic traders at time $t$ of a total number of \Chartists chartists.
They are assumed to change their sentiment based on the (extensive) transition rates $\pi^+ = a + bn$ and $\pi^- = a + b(N-n)$, where parameter $a$ indicates idiosyncratic switches and $b$ measures the herding intensity.
The resulting sentiment dynamics is approximated by the following Langevin equation with drift component $A(x) = -2ax$ and diffusion term $D(x) = 2b(1-x^2_t) + \nicefrac{4a}{N}$ which gives
\begin{equation} \elabel{FokkerPlanckApprox}
	\dif x_t = A(x) \dt + D(x) \dif W_t = -2ax_t \dt + \sqrt{2b(1-x^2_t)} \dif W_t ~.
\end{equation}
\eref{FokkerPlanckApprox} can be discretized with $\Delta t = 1$:
\begin{equation}
	\Delta x = x_{t+\Delta t} - x_t = -2ax_t+\sqrt{2b\left(1-x_t^2\right)}~\epsilon_t
\end{equation}
with $\epsilon\sim \ND{0}{1}$.
The distribution of the sentiment index is known to be bimodal for $a<b$ and unimodal for $a>b$.

Price dynamics are governed by a standard Walrasian adjustment mechanism, depending on total excess demand of both trading groups:
\begin{equation}
\begin{split}
p_{t+1}	&= p_t + \beta\left(\DemandFt + \DemandCt \right) \\
 				&= p_t + \beta\left[ \FundisT \Vft \left(p^{\star}_t-p_t\right)+ \ChartistsT \Vct x_t \right] ~,
\end{split}
\end{equation}
where $\beta$ is the assumed price adjustment speed.
With instantaneous market clearing, \ie $\beta\to\infty$, and setting $\nicefrac{\Chartists\Vc}{\Fundis \Vf}=1$, we get the following evolution of returns:
\begin{equation}\elabel{priceDynALW}
\begin{split}  
r_{t+1} &= p_{t+1}-p_t \\
        &= \SigmaF \epsilon_t + \left( x_t-x_{t-1} \right) ~.
\end{split}
\end{equation}
Finally, the parameter vector to be estimated for the \ALWmodel contains three items $\paramVecALW = \left( a,b,\SigmaF \right)^\top$.
In accordance with the literature we use for all our simulation experiments the following true model parameters
$\paramVecTrue = \left( 0.3, 1.4, 30 \right)^\top$.%
\footnote{
See \textcite{GhonghadzeLux2016,ChenLux2018}.
The (pseudo-)true parameter setting for the \ALWmodel that is used in this paper represents the bimodal case of the underlying sentiment index given that $b > a$.
Since \textcite{ChenLux2018} have shown that the bimodal case gives the best fit to empirical data, we will only focus on this model scenario.
Note that for better readability parameters of the \ALWmodel are always multiplied by $10^3$ throughout the paper.
}


\subsubsection{FW model}\sseclabel{FWmodelDescription}
\textcite{FrankeWesterhoff2012} propose an entire model zoo for which they run a model contest.
Here, we will only consider their top performing DCA-HPM model version (discrete choice with herding, predisposition and price misalignment).
\textcite{FrankeWesterhoff2012} apply a market maker model that considers as well two types of speculators: chartist and fundamentalists, whose fractions are denoted by \FracChart and \FracFundis, respectively.
The evolution of the log prices is then determined by
\begin{equation}
	p_{t+1} = p_t + \mu\left(\FracFundisT \DemandFt + \FracChartT \DemandCt \right) ~,
\end{equation}
with $\mu$ reflecting the speed of price adjustment.
Excess demand of chartists \DemandC and fundamentalists \DemandF is given by
\begin{equation}
\begin{split}
\DemandFt & = \phi\left(p^{\star}-p_t\right) + \epsFundisT \\
\DemandCt & = \chi\left(p_t-p_{t-1}\right) + \epsChartT ~,
\end{split}
\end{equation}
with $\epsFundis \sim \ND{0}{\SigmaF^2}$ and $\epsChart \sim \ND{0}{\SigmaC^2}$. Parameters $\chi$ and $\phi$ are always positive and indicate the strength of reaction. The switching between both trading strategies is governed by a fitness function measuring the attractiveness $a_t$ of fundamentalism over chartism:
\begin{equation}
	a_t=
	\underbrace{
		\alpha_n\left( \FracFundisT - \FracChartT \right)
	}_{\text{herding intensity}} 
	+ \alpha_0 +
	\underbrace{
		\alpha_p\left(p_t-p^{\star}\right)^2
	}_{\text{misalignment}} 
	~,
\end{equation}
where $\alpha_0$ represents a constant idiosyncratic predisposition for one of the two trading strategies, $\alpha_n$ relates to herding intensity and $\alpha_p$ accounts for price misalignments from the fundamental value $p^\star$. Note that while $\alpha_n$ and $\alpha_p$ are always strictly positive, parameter $\alpha_0$ might be negative as well.
The current market shares of fundamentalists \FracFundisT and chartists \FracChartT at time $t$ are then updated according to the following discrete choice approach:
\begin{equation}
\begin{split}
\FracFundisT & = \frac{1}{1+\exp\left(-\beta a_{t-1}\right)}\\
\FracChartT  & = 1-\FracFundisT
\end{split}
\end{equation}
As in \textcite[1199]{FrankeWesterhoff2012} we set the intensity of choice parameter to $\beta=1$, the speed of price adjustment parameter to $\mu=0.01$ and the fundamental price to $p^*=0$.
Finally, the parameter vector to be estimated for the \FWmodel contains a total of seven parameters $\paramVecFW = \left( \phi,\chi,\alpha_0,\alpha_n,\alpha_p,\SigmaF,\SigmaC \right)^\top$.
In accordance with \textcite{FrankeWesterhoff2012} we use for all our simulation experiments the following true model parameters
$\paramVecTrue = \left(0.12,1.5,-0.336,1.839,19.671,0.708,2.147\right)^\top$.

To sum up, the \ALWmodel by \textcite{AlfaranoEtAl2008} is a small-scale model which can be reduced to only three estimation parameters $\paramVecALW = \left(a,b,\sigma_f\right)^\top$.
The \FWmodel by \textcite{FrankeWesterhoff2012} is more complex with a total of seven parameters $\paramVecFW= \left( \phi,\chi,\alpha_0,\alpha_n,\alpha_p,\SigmaF,\SigmaC \right)^\top$. Sample simulation runs for both model dynamics can be found in appendix \ref{appendix:AppendixSampleSimulationRuns}.

\subsection{Moment functions} \sseclabel{Moments}
The following subsection builds on \sref{Method} and explains the motivation behind the choice of our set of moment functions.
As explained in \ssref{FABMs}, we focus on two agent-based asset pricing models which replicate many of the stylized facts of financial markets.
The vector of moment functions \momVec should contain a set of reasonable summary statistics that (partly) capture these stylized facts expressed as observables and thus measurable statistical quantities.
A necessary condition for the choice of moment functions is given by the order condition, \ie for estimations with more than one model parameter the number of moment functions \MomentsVecSize needs to be greater or equal than the number of model parameters or the cardinality of the parameter vector \paramVec.
Thus, the order condition provides a lower bound for the number of moment conditions.
Theoretically, the (full) rank condition is a sufficient condition for identification assuring that only moment functions without linear dependence are included.
While the order condition is easy to meet, the rank condition is barely testable for most (F)ABMs given their non-linearity and the lack of analytical closed-form expressions.


\begin{table}[h!]
\caption{\textbf{Moment functions}. We use a set of eighteen moment functions to cover the most important stylized facts. Therefore we include the returns' volatility defined as the mean value of absolute returns, unconditional second and fourth return moments, Hill estimators of the power law tail index for the top 2.5\% and top 5\%, $\alpha_{2.5}$ and $\alpha_{5.0}$, the first order autocorrelation coefficient of raw returns and the autocorrelation coefficients of absolute and squared returns for lags $1,5,10,25,50$ and 100.}
\tlabel{MomentFunctions}
\centering
\footnotesize
\begin{tabular}{lcl}
\addlinespace
\toprule
Moment function & Notation & Statistic \\ 
\midrule
expectation value										& $m_1$			& \EV{|r_t|} \\
\addlinespace
variance 														& $m_2$ 		& \EV{r_t^2} \\
\addlinespace
kurtosis 														& $m_3$ 		& \EV{r_t^4} \\
\addlinespace
power law tail exponent top 2.5\%		& $m_4$			& $\alpha_{2.5}$  \\
\addlinespace
power law tail exponent top 5\%			& $m_5$			& $\alpha_{5.0}$  \\
\addlinespace
AC raw returns lag 1								& $m_6$			& \EV{r_tr_{t-1}} \\
\addlinespace
AC absolute returns lag 1						& $m_7$			& \EV{|r_t||r_{t-1}|} \\
\addlinespace
AC squared returns lag 1						& $m_8$			& \EV{r_t^2 r_{t-1}^2} \\
\addlinespace
AC absolute returns lag 5						& $m_9$			& \EV{|r_t||r_{t-5}|}	\\
\addlinespace
AC squared returns lag 5						& $m_{10}$	& \EV{r_t^2 r_{t-5}^2} \\
\addlinespace
AC absolute returns lag 10					&	$m_{11}$	& \EV{|r_t||r_{t-10}|}	\\
\addlinespace
AC squared returns lag 10						& $m_{12}$	& \EV{r_t^2 r_{t-10}^2} \\
\addlinespace
AC absolute returns lag 25					&	$m_{13}$	& \EV{|r_t||r_{t-25}|}	\\
\addlinespace
AC squared returns lag 25						& $m_{14}$	& \EV{r_t^2 r_{t-25}^2} \\
\addlinespace
AC absolute returns lag 50					&	$m_{15}$	& \EV{|r_t||r_{t-50}|}	\\
\addlinespace
AC squared returns lag 50						& $m_{16}$	& \EV{r_t^2 r_{t-50}^2} \\
\addlinespace
AC absolute returns lag 100					&	$m_{17}$	& \EV{|r_t||r_{t-100}|} \\
\addlinespace
AC squared returns lag 100					& $m_{18}$	& \EV{r_t^2 r_{t-100}^2} \\
\bottomrule
\end{tabular}
\end{table}

Besides these restrictions the number and choice of moment conditions is not strictly limited which may render it arbitrary.
However, this is not a weakness \perse, because SMM is designed to capture complex patterns which likely escape any single moment condition.
Thus, if empirical data show complex patterns that are hard to squeeze into a single metric, it can be necessary to wrap the patterns in more than a single moment condition.
The stylized facts of financial markets show such complex patterns like bubbles and crashes, excess volatility, heavy-tailed return distributions, absence of autocorrelation in raw returns or slow decay in volatility.
Especially the statistical pattern of slow non-linear decay in autocorrelation over many lags requires more than one moment condition.
This justifies the comparatively large size of our vector of moment functions with $\MomentsVecSize = 18$
listed in \Tref{MomentFunctions}.
For the SMM approach this is a common size of the moment vector, see \eg \textcite[722]{ChenLux2018} who analyse the \ALWmodel with up to 15 moment functions or \textcite[930]{RugeMurcia2012} who uses 16 moment functions for the estimation of a macroeconomic DSGE model.


As the first two moment functions we use the mean of absolute returns and the variance of the return series.
As mentioned in \sref{Method}, ordinary moments in the sense of \eref{MomentsMeasure} for HTDs do only exist for orders lower than the tail index.
For example, if a return series has a tail index of $\alpha = 3$, only the first and second moment exist.
This means that sample estimates of the third or higher moments will never converge with increasing sample size.
Hence, such moments might not be suitable as moment functions for estimation purposes.
We yet consider the kurtosis of the return series as one moment function and additionally include the Hill tail index estimates at \SI{2.5}{\percent} and \SI{5.0}{\%}.
The autocorrelation of raw returns at lag 1 checks for the absence of serial correlation in returns.
Finally, in order to capture the slow decay of volatility we include autocorrelations of absolute and squared returns for the lags $1,5,10,25,50$ and $100$.


\Tref{MomentFunctions} offers an extensive yet non-comprehensive list of possible summary statistics considered for the estimation of financial market models.
We abstain from modelling moment functions as further derived processes, like a GARCH$(1,1)$, firstly because of the additional computational cost associated with the estimation of the GARCH parameters. Secondly and more importantly, they introduce additional uncertainty leading to possible biases in the estimates.

\afterpage{
\clearpage							
\begin{landscape}
\centering 
\begin{table}
\caption{True (theoretical) values of moment functions. The table lists mean value, variance and p-value of Kolmogorov-Smirnov statistic for the ALW and FW model for all 18 moment functions listed in \tref{MomentFunctions} over $T_1=\num{10000}$, $T_2=\num{100000}$, $T_3=\num{1000000}$ and $T_4=\num{1000000}$. Simulations are run over $\MCSize = \num{5000}$ repetitions.}
\tlabel{EnsAverages}
\resizebox{\columnwidth}{!}{%
\begin{tabular}{llrrrrrrrrrrrrrrrrrrrrrrrrrrr}
\addlinespace
\toprule
& & \multicolumn{3}{c}{$\Moment_1$}	& \multicolumn{3}{c}{$\Moment_2$}	& \multicolumn{3}{c}{$\Moment_3$} & \multicolumn{3}{c}{$\Moment_4$}	& \multicolumn{3}{c}{$\Moment_5$}	& \multicolumn{3}{c}{$\Moment_6$} & \multicolumn{3}{c}{$\Moment_7$}	& \multicolumn{3}{c}{$\Moment_8$}	& \multicolumn{3}{c}{$\Moment_9$}\\
\cmidrule(lr){3-5} \cmidrule(lr){6-8} \cmidrule(lr){9-11} \cmidrule(lr){12-14} \cmidrule(lr){15-17} \cmidrule(lr){18-20} \cmidrule(lr){21-23} \cmidrule(lr){24-26} \cmidrule(lr){27-29}
    &    & mean   & var    & p-value & mean   & var    & p-value & mean   & var    & p-value & mean   & var    & p-value & mean   & var    & p-value & mean    & var    & p-value & mean   & var    & p-value & mean   & var    & p-value & mean   & var    & p-value \\
\cmidrule(lr){3-29}
\multirow{4}[1]{*}{\textbf{ALW}} & T1 & 0.0332 & 0.0    & 0.0365  & 0.0019 & 0.0    & 0.0153  & 0.8199 & 0.0184 & 0.578   & 5.7092 & 0.1543 & 0.0038  & 4.7256 & 0.0543 & 0.0083  & -0.0    & 0.0001 & 0.7456  & 0.0993 & 0.0002 & 0.8175  & 0.0953 & 0.0002 & 0.6847  & 0.0973 & 0.0002 & 0.7464  \\
    & T2 & 0.0332 & 0.0    & 0.668   & 0.0019 & 0.0    & 0.6258  & 0.8529 & 0.0019 & 0.7856  & 5.6635 & 0.0159 & 0.7591  & 4.6839 & 0.0057 & 0.7714  & -0.0002 & 0.0    & 0.467   & 0.1034 & 0.0    & 0.9827  & 0.0983 & 0.0    & 0.6949  & 0.1015 & 0.0    & 0.3522  \\
    & T3 & 0.0332 & 0.0    & 0.4123  & 0.0019 & 0.0    & 0.7276  & 0.8563 & 0.0002 & 0.9226  & 5.6597 & 0.0016 & 0.8183  & 4.6804 & 0.0006 & 0.2509  & -0.0002 & 0.0    & 0.9348  & 0.1038 & 0.0    & 0.8736  & 0.0986 & 0.0    & 0.979   & 0.1019 & 0.0    & 0.9412  \\
    & T4 & 0.0332 & 0.0    & 0.7663  & 0.0019 & 0.0    & 0.6951  & 0.8563 & 0.0    & 0.6985  & 5.6595 & 0.0002 & 0.9216  & 4.6799 & 0.0001 & 0.7177  & -0.0002 & 0.0    & 0.9835  & 0.1038 & 0.0    & 0.9149  & 0.0986 & 0.0    & 0.9088  & 0.1019 & 0.0    & 0.834   \\
\cmidrule(lr){2-29}
    &    & \multicolumn{3}{c}{$\Moment_{10}$}	& \multicolumn{3}{c}{$\Moment_{11}$}	& \multicolumn{3}{c}{$\Moment_{12}$}	& \multicolumn{3}{c}{$\Moment_{13}$} & \multicolumn{3}{c}{$\Moment_{14}$}	& \multicolumn{3}{c}{$\Moment_{15}$}	& \multicolumn{3}{c}{$\Moment_{16}$}	& \multicolumn{3}{c}{$\Moment_{17}$}	& \multicolumn{3}{c}{$\Moment_{18}$}\\
\cmidrule(lr){3-5} \cmidrule(lr){6-8} \cmidrule(lr){9-11} \cmidrule(lr){12-14} \cmidrule(lr){15-17} \cmidrule(lr){18-20} \cmidrule(lr){21-23} \cmidrule(lr){24-26} \cmidrule(lr){27-29}
    &    & mean   & var    & p-value & mean   & var    & p-value & mean   & var    & p-value & mean   & var    & p-value & mean   & var    & p-value & mean    & var    & p-value & mean   & var    & p-value & mean   & var    & p-value & mean   & var    & p-value \\
\cmidrule(lr){3-29}
\multirow{4}{*}{\textbf{ALW}} & T1 & 0.0935 & 0.0003 & 0.7791  & 0.0951 & 0.0002 & 0.5688  & 0.0915 & 0.0002 & 0.2882  & 0.0884 & 0.0002 & 0.9002  & 0.0851 & 0.0002 & 0.2923  & 0.0778  & 0.0002 & 0.8222  & 0.0749 & 0.0002 & 0.4867  & 0.0607 & 0.0002 & 0.9124  & 0.0586 & 0.0002 & 0.2626  \\
    & T2 & 0.0966 & 0.0    & 0.768   & 0.0991 & 0.0    & 0.9564  & 0.0944 & 0.0    & 0.5326  & 0.0924 & 0.0    & 0.9785  & 0.0882 & 0.0    & 0.9196  & 0.0824  & 0.0    & 0.6354  & 0.0787 & 0.0    & 0.8017  & 0.0657 & 0.0    & 0.8953  & 0.063  & 0.0    & 0.5687  \\
    & T3 & 0.0969 & 0.0    & 0.786   & 0.0996 & 0.0    & 0.9963  & 0.0947 & 0.0    & 0.9349  & 0.0929 & 0.0    & 0.347   & 0.0886 & 0.0    & 0.4198  & 0.083   & 0.0    & 0.8863  & 0.0793 & 0.0    & 0.4013  & 0.0662 & 0.0    & 0.9985  & 0.0634 & 0.0    & 0.9486  \\
    & T4 & 0.0969 & 0.0    & 0.4051  & 0.0996 & 0.0    & 0.6847  & 0.0947 & 0.0    & 0.7574  & 0.093  & 0.0    & 0.947   & 0.0886 & 0.0    & 0.8556  & 0.083   & 0.0    & 0.9503  & 0.0793 & 0.0    & 0.7092  & 0.0662 & 0.0    & 0.8786  & 0.0635 & 0.0    & 0.7422  \\
\addlinespace
\midrule
\addlinespace
& & \multicolumn{3}{c}{$\Moment_1$}	& \multicolumn{3}{c}{$\Moment_2$}	& \multicolumn{3}{c}{$\Moment_3$} & \multicolumn{3}{c}{$\Moment_4$}	& \multicolumn{3}{c}{$\Moment_5$}	& \multicolumn{3}{c}{$\Moment_6$} & \multicolumn{3}{c}{$\Moment_7$}	& \multicolumn{3}{c}{$\Moment_8$}	& \multicolumn{3}{c}{$\Moment_9$}\\
\cmidrule(lr){3-5} \cmidrule(lr){6-8} \cmidrule(lr){9-11} \cmidrule(lr){12-14} \cmidrule(lr){15-17} \cmidrule(lr){18-20} \cmidrule(lr){21-23} \cmidrule(lr){24-26} \cmidrule(lr){27-29}
    &    & mean   & var    & p-value & mean   & var    & p-value & mean   & var    & p-value & mean   & var    & p-value & mean   & var    & p-value & mean    & var    & p-value & mean   & var    & p-value & mean   & var    & p-value & mean   & var    & p-value \\
\cmidrule(lr){3-29}
\multirow{4}{*}{\textbf{FW}} & T1 & 0.0071 & 0.0    & 0.0584  & 0.0001 & 0.0    & 0.0249  & 2.3495 & 0.0781 & 0.0021  & 4.3731 & 0.1543 & 0.0001  & 3.5976 & 0.0401 & 0.0     & 0.0073  & 0.0002 & 0.2485  & 0.1881 & 0.0004 & 0.0     & 0.1796 & 0.0004 & 0.5184  & 0.1839 & 0.0004 & 0.0     \\
    & T2 & 0.0071 & 0.0    & 0.2857  & 0.0001 & 0.0    & 0.3108  & 2.4246 & 0.0074 & 0.9997  & 4.3516 & 0.0175 & 0.139   & 3.5312 & 0.0039 & 0.0056  & 0.0072  & 0.0    & 0.4614  & 0.1937 & 0.0    & 0.1256  & 0.1824 & 0.0    & 0.7128  & 0.1893 & 0.0    & 0.1071  \\
    & T3 & 0.0071 & 0.0    & 0.9359  & 0.0001 & 0.0    & 0.5581  & 2.4302 & 0.0007 & 0.6599  & 4.3502 & 0.0018 & 0.5474  & 3.5254 & 0.0004 & 0.8748  & 0.0072  & 0.0    & 0.6108  & 0.1942 & 0.0    & 0.1158  & 0.1826 & 0.0    & 0.8993  & 0.1898 & 0.0    & 0.7911  \\
    & T4 & 0.0071 & 0.0    & 0.9987  & 0.0001 & 0.0    & 0.9397  & 2.4314 & 0.0001 & 0.9388  & 4.3505 & 0.0002 & 0.4357  & 3.5249 & 0.0    & 0.9837  & 0.0072  & 0.0    & 0.5315  & 0.1943 & 0.0    & 0.6216  & 0.1826 & 0.0    & 0.9461  & 0.1899 & 0.0    & 0.4562  \\
\cmidrule(lr){2-29}
    &    & \multicolumn{3}{c}{$\Moment_{10}$}	& \multicolumn{3}{c}{$\Moment_{11}$}	& \multicolumn{3}{c}{$\Moment_{12}$}	& \multicolumn{3}{c}{$\Moment_{13}$} & \multicolumn{3}{c}{$\Moment_{14}$}	& \multicolumn{3}{c}{$\Moment_{15}$}	& \multicolumn{3}{c}{$\Moment_{16}$}	& \multicolumn{3}{c}{$\Moment_{17}$}	& \multicolumn{3}{c}{$\Moment_{18}$}\\
\cmidrule(lr){3-5} \cmidrule(lr){6-8} \cmidrule(lr){9-11} \cmidrule(lr){12-14} \cmidrule(lr){15-17} \cmidrule(lr){18-20} \cmidrule(lr){21-23} \cmidrule(lr){24-26} \cmidrule(lr){27-29}
    &    & mean   & var    & p-value & mean   & var    & p-value & mean   & var    & p-value & mean   & var    & p-value & mean   & var    & p-value & mean    & var    & p-value & mean   & var    & p-value & mean   & var    & p-value & mean   & var    & p-value \\
\cmidrule(lr){3-29}
\multirow{4}{*}{\textbf{FW}} & T1 & 0.1742 & 0.0004 & 0.3067  & 0.1764 & 0.0004 & 0.0     & 0.1661 & 0.0004 & 0.0672  & 0.1497 & 0.0005 & 0.0     & 0.1388 & 0.0004 & 0.0853  & 0.1112  & 0.0005 & 0.0029  & 0.1013 & 0.0004 & 0.8188  & 0.0641 & 0.0005 & 0.8457  & 0.057  & 0.0005 & 0.5279  \\
    & T2 & 0.1766 & 0.0    & 0.4711  & 0.1819 & 0.0    & 0.1044  & 0.1689 & 0.0    & 0.9349  & 0.1553 & 0.0    & 0.1422  & 0.1423 & 0.0    & 0.7186  & 0.117   & 0.0    & 0.9478  & 0.1058 & 0.0    & 0.8696  & 0.0698 & 0.0001 & 0.3663  & 0.0619 & 0.0001 & 0.0873  \\
    & T3 & 0.177  & 0.0    & 0.6711  & 0.1824 & 0.0    & 0.3943  & 0.1691 & 0.0    & 0.9735  & 0.1559 & 0.0    & 0.7974  & 0.1428 & 0.0    & 0.9909  & 0.1176  & 0.0    & 0.788   & 0.1063 & 0.0    & 0.9977  & 0.0704 & 0.0    & 0.8684  & 0.0624 & 0.0    & 0.8808  \\
    & T4 & 0.177  & 0.0    & 0.9552  & 0.1825 & 0.0    & 0.9178  & 0.1692 & 0.0    & 0.9023  & 0.156  & 0.0    & 0.4667  & 0.1428 & 0.0    & 0.8939  & 0.1177  & 0.0    & 0.5447  & 0.1064 & 0.0    & 0.6114  & 0.0705 & 0.0    & 0.3633  & 0.0625 & 0.0    & 0.7818 \\
\bottomrule
\end{tabular}
}
\end{table}
\end{landscape}
}

\subsubsection{True moment vector}\ssseclabel{TrueMomentVector}

In order to apply any moment estimator, it is important that simulated moments (asymptotically) equal their true values.
This means that true values exist and convergence happens with an increasing number of observations (either over time \Time and/or the ensemble \EnsembleSize).
Otherwise, meaningful inference is not possible.
Hence, the very first step of our analysis is to numerically check if the moment functions do exist, \ie if they quickly converge to identical values for some finite number of observations.
Since the mDGP for most (F)ABMs is not available in closed form, the ensemble average \EA{X} is not readily at hand.
Living in a pre-asymptotic numerical simulation, sufficient convergence to the theoretical average needs to take place rather quickly for finite observations.
Recall that we talked about the conditions for the existence of higher moments in the previous section.
To find the moments' true values, we run simulations over different time lengths \Time with $\Time_1=\num{10 000}$, $\Time_2=\num{100 000}$, $\Time_3=\num{1 000 000}$ and $\Time_4=\num{10000000}$. 
We run each scenario over $\MCSize = \num{5000}$ replications.
Results are presented in \tref{EnsAverages}.
For each moment function, we report mean, variance and $p$-value of a Kolmogorov-Smirnov test statistic.

We observe convergence behaviour for all moment functions for both the \ALWmodel and \FWmodel for an increased number of observations.
In fact, the variance decreases consistently for all considered moment functions and models.
Given that the variance is low for all considered moment functions, we can assume that the long-run estimates are informative for the estimation.
Therefore, we will consider the estimates of the longest time length $T_4$ to compute the inverse of the long-run covariance matrix as the optimal weighting matrix in our SMM estimator.
High $p$-values for the KS-test also suggest normally distributed samples of the moment estimates.
Given that the true values exit and converge, we can move on to the next step which is to test if the chosen moment functions are suited for identification of the model parameters.

\FloatBarrier
\subsubsection{Pre-asymptotic properties of moment functions} \sseclabel{PreAsymptoticsMomentFunctions}
In this subsection we are concerned with the pre-asymptotic properties of our moment functions and their identification power.
We want to check if the mapping between model parameters and model output in terms of moment functions is unique. Only a one-to-one mapping works as a sufficient condition (\ie full rank condition) for identification.
For the one parameter case, the identification issue is quite obvious.
A sufficient condition for identification would be a strictly monotone relationship between model parameter and moment function.
For higher dimensional models, this is less trivial.
In fact, as stated earlier, the rank condition is barely directly testable for ABMs.

Nevertheless, running sensitivity analyses allows us to study how strongly moment functions react towards changes in model parameters.\footnote{Since not all moment functions are equally informative, another way to identify relevant moment functions is in terms of statistical efficiency. \textcite{GallantTauchen1996} suggest using scores of an auxiliary model as moment functions in a generalized method of moment (GMM) estimator in order to reduce loss of efficiency due to uninformative moments.
Hence, by searching for the statistically most informative moments, such an approach tries to reach the efficiency of maximum-likelihood.
Auxiliary models are considered to be especially attractive when the number of moment functions is rather limited.}
Therefore, we break down the model's dimensionality and study how sensitive our moment functions are to variations of the model parameters.

\begin{figure}[hp]
\begin{center}
 		\includegraphics[width=\textwidth]{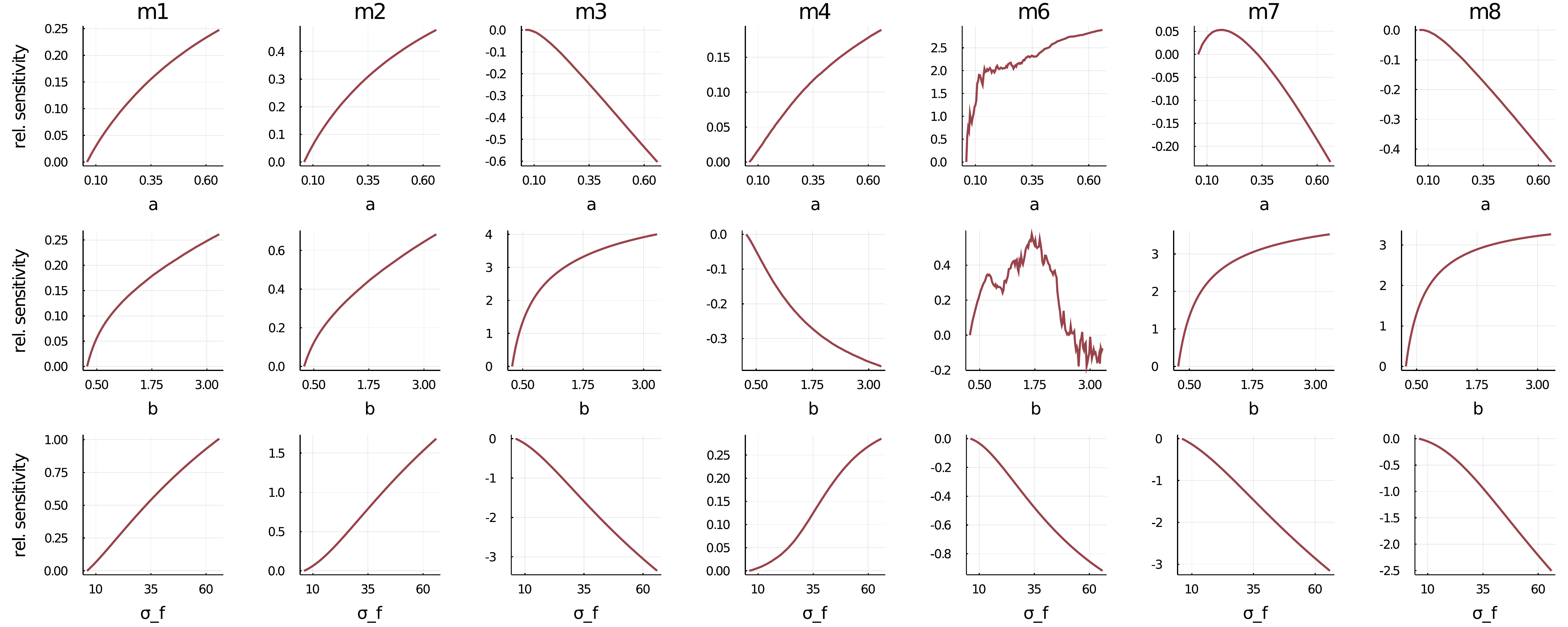} 	
	\caption{\textbf{Sensitivity of the moment functions to model parameters in the \ALWmodel.}
	Model parameters are varied on horizontal axis, moment response is on the vertical axis.
	The top panel shows selected moment function responses for parameter $a$, the middle panel for parameter $b$ and the bottom panel for parameter \SigmaF.
	Simulations are run over $\MCSize = 100$ Monte Carlo replications with $\Time = \num{20 000}$ and $\EnsembleSize = 50$.}
\flabel{MomentParameterSensitivityALW}
\end{center}
\end{figure}

For the \ALWmodel, we sample 151 equidistant points over a large range of parameter variations: $a \in [0.06, 0.66]$, $b \in [0.28, 3.28]$ and $\sigma_f \in [6.0, 66.0]$. We plot the relative responsiveness of moments $m_1$, $m_2$, $m_3$, $m_4$, $m_6$, $m_7$ and $m_8$ for the \ALWmodel in \fref{MomentParameterSensitivityALW}.\footnote{Moment functions $m_9$ to $m_18$ for the autocorrelations of absolute and squared returns at higher lags show qualitatively the same functional behaviour as $m_7$ and $m_8$, respectively.}
We find non-linear relationships for all moment-parameter pairs.
Moment function $m_6$ (autocorrelation of raw returns at lag 1) shows highly non-smooth behaviour towards changes in model parameters $a$ and $b$ of the \ALWmodel.
Regarding the strength of reactions, moment functions react most sensitive to parameter $\sigma_f$. This is in line with earlier findings by \textcite{ChenLux2018}.
In fact, the moment functions are least sensitive towards varying values of parameter $a$.
\afterpage{
\clearpage
\begin{landscape}
\begin{figure}[h]
\begin{center}
 		\includegraphics[height=.8\textwidth]{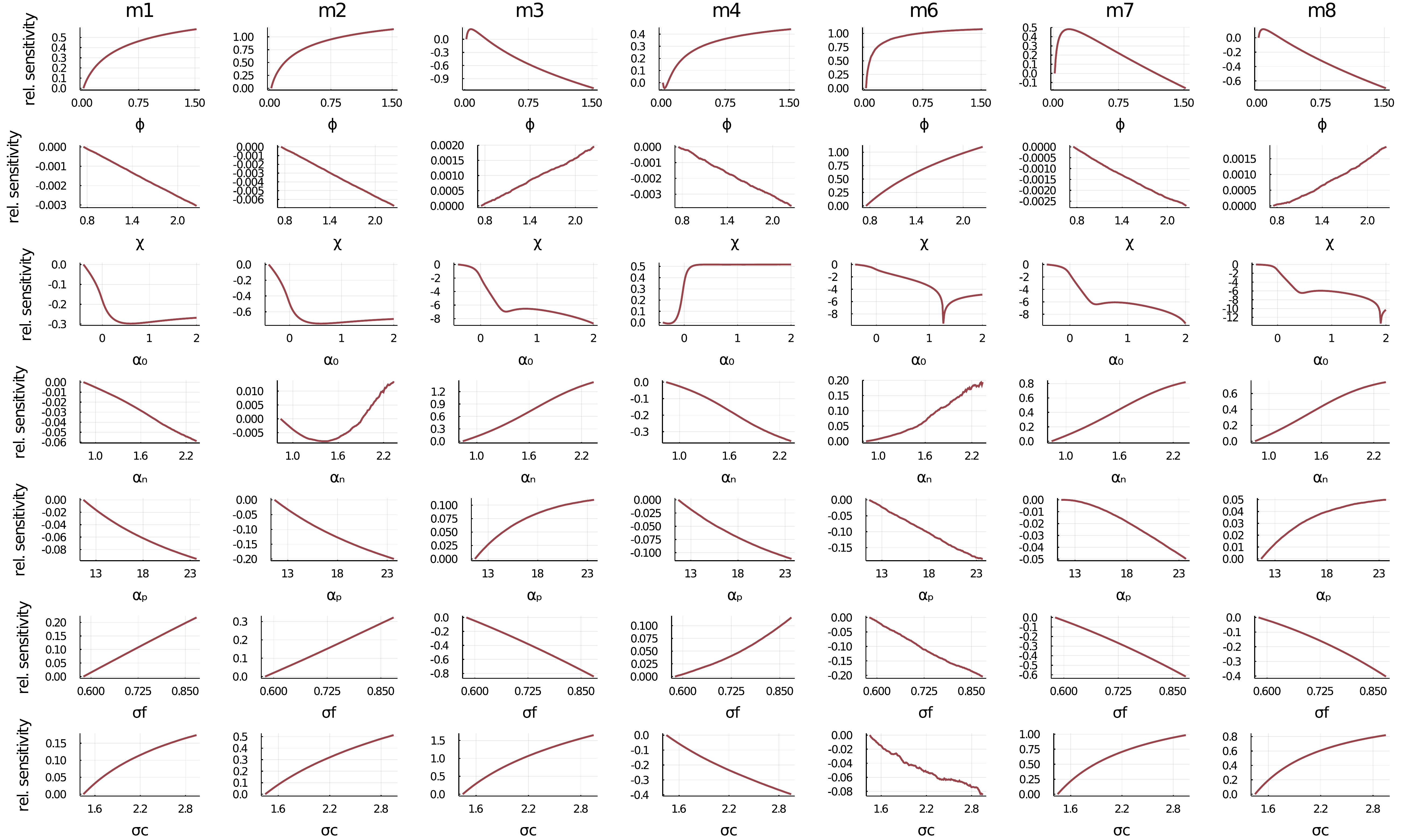}
	\caption{\textbf{Sensitivity of the moments to model parameters in the \FWmodel.}
	Model parameters are varied on horizontal axis, moment response is on the vertical axis.
	The panels show selected moment responses for the seven parameters of the \FWmodel: $\phi, \chi, \alpha_0, \alpha_n, \alpha_p,$ \SigmaF and \SigmaC.
	Simulations are run over $\MCSize = 100$ Monte Carlo replications with $\Time = \num{20 000}$ and $\EnsembleSize = 50$.}
\flabel{MomentParameterSensitivityFW}
\end{center}
\end{figure}
\end{landscape}
\clearpage
}

For the \FWmodel, we also run simulation for 151 equidistant points over a broad range of parameter variations:
$\phi \in [0.02, 1.52]$,
$\chi \in [0.75, 2.25]$,
$\alpha_0 \in [-0.4, 2.0]$,
$\alpha_n \in [0.839, 2.339]$,
$\alpha_p \in [11.671, 23.671]$,
$\SigmaF \in [0.58, 0.88]$ and
$\SigmaC \in [1.447, 2.947]$.
Results are shown in \fref{MomentParameterSensitivityFW}. We find again highly non-linear functional relationships for almost all moment-parameter combinations. We further see parameter $\alpha_0$ as probably hard to identify correctly given the wild and untamed nature of the moment function responses. Additionally, moment functions react comparatively insensitive towards parameter $\chi$ suggesting that it might trigger identification problems, too.

To conclude, sensitivity analyses reveal that most functional relationships are non-linear which might result in (small) biases of the estimates.
As mentioned in \textcite{GrazziniEtAl2012,GrazziniRichiardi2015}, if a moment function happens to be non-linear, there will be a small bias with the direction of the bias depending on moment functions' derivatives.
Thereby concave (convex) moment functions lead to an upward (downward) bias.
Such biases can be reduced in different ways: (i) given the analytical expression of the moment function, however, this is unknown for most ABMs, (ii) through monotonic transformations and (iii) through increased number of observations.
The latter brings us to the focus of our study.

Recall that biases decrease for increasing number of observations since the simulated moment functions converge to their true theoretical value.
Hence, the next step is to take a closer look at the convergence speed of the moment functions.
The goal here is twofold.
First, we want to check if an ensemble \EnsembleSize of simulation runs converges to the exact same value as the corresponding long-run realization, \ie how strong the effect of broken ergodicity is.
Second, we aim at evaluating the speed of convergence.
\begin{figure}[h!]
  \centering
	\subcaptionbox{\ALWmodel.\flabel{MomentConvergenceALW}}{\includegraphics[width=.49\textwidth]{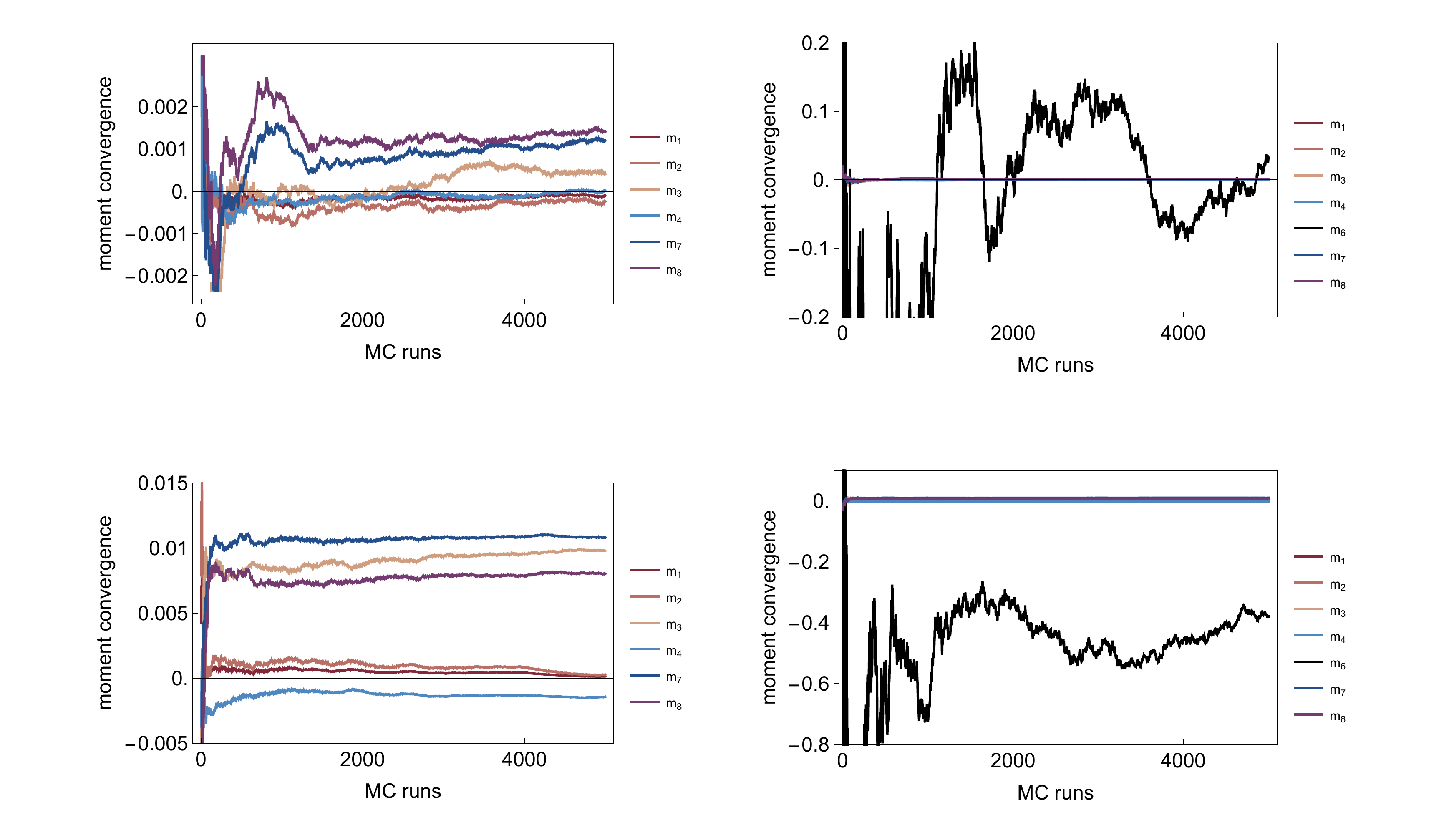}}
	\subcaptionbox{\FWmodel.\flabel{MomentConvergenceFW}}{\includegraphics[width=.49\textwidth]{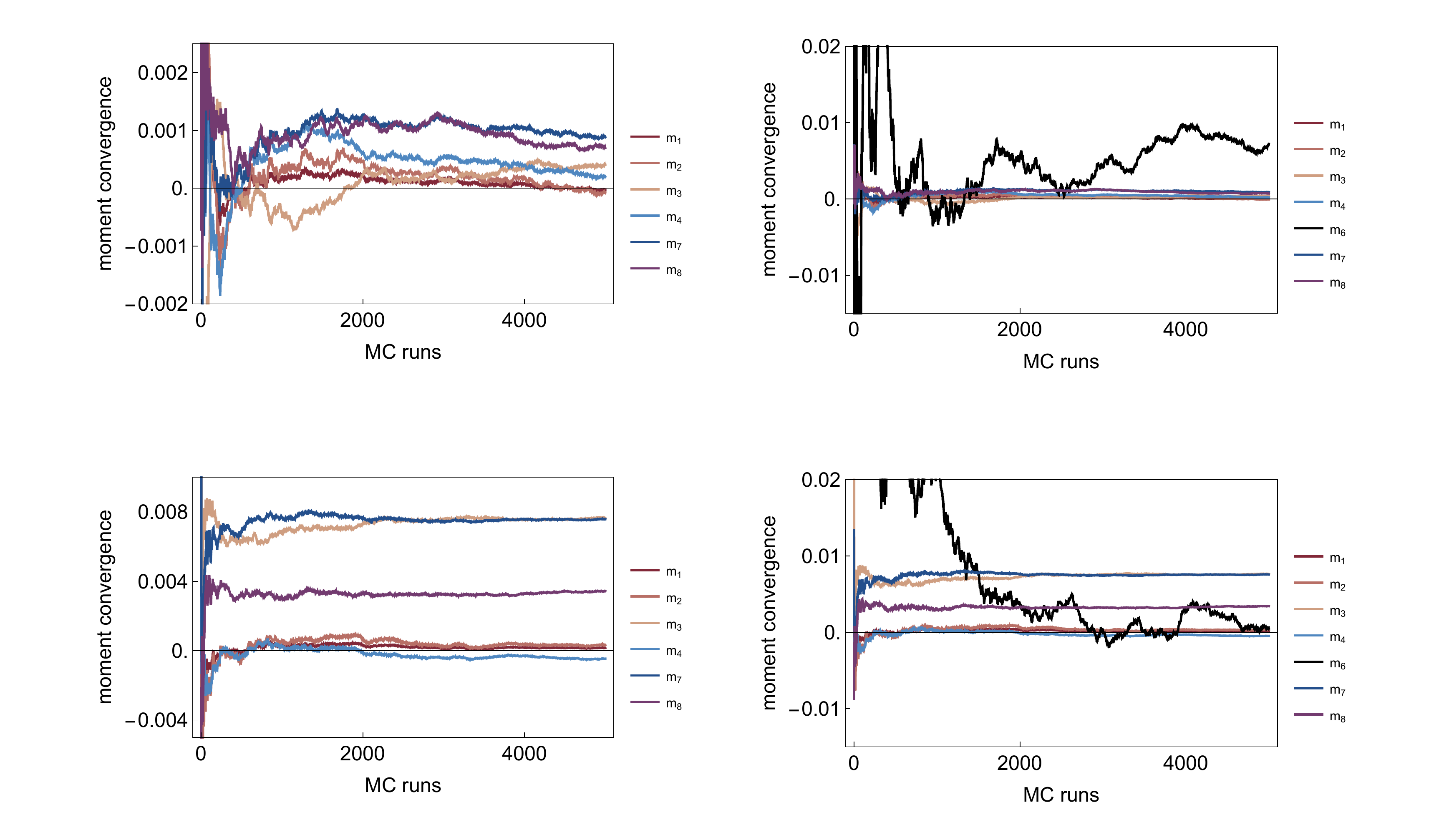}}
\caption{\textbf{Convergence behaviour for selected moment functions.}
Plots reveal the relative deviations of moment functions from their true values for the \ALWmodel in \fref{MomentConvergenceALW} and for the \FWmodel in \fref{MomentConvergenceFW}.
The horizontal axis is representing the number of Monte Carlo runs ($\MCSize = \num{5 000}$).}
\flabel{MomentConvergence}
\end{figure}
For this experiment, we run simulations over $\MCSize = \num{5000}$ repetitions for two different scenarios:
(i) $\Time = \num{400000}$, $\EnsembleSize = 1$ and
(ii) $\Time = \num{40000}$, $\EnsembleSize = 10$.
In \fref{MomentConvergence} we plot again the convergence behaviour for the moment functions $m_1$, $m_2$, $m_3$, $m_4$, $m_6$, $m_7$ and $m_8$ defined as relative deviations from their true values, \ie their long-run estimates reported in \sssref{TrueMomentVector}.
The top and lower panels of \fref{MomentConvergenceALW} and \fref{MomentConvergenceFW} show results for scenario (i) and (ii), respectively.
The left (right) panels refer to the selected moment function including $m_6$ (excluding $m_6$).
We find for both the \ALWmodel and \FWmodel that moment functions $m_1$, $m_2$ and $m_4$ closely converge towards their true values, while $m_3$, $m_7$ and $m_8$ show more or less pronounced and persistent biases.
Regarding the speed of convergence, we observe long transition phases for scenario (i) and comparatively fast adjustments for scenario (ii).
Yet, deviations are more severe here, too.
Looking at the right panels including moment function $m_6$ which is the autocorrelation of raw returns at lag 1, we find very volatile and probably non-convergent behaviour for both models.
This suggests that moment $m_6$ might not be suited for identification of the given model outcomes.

\FloatBarrier
\subsection{Properties of the objective function}\sseclabel{NumericalExperiments}

Next, we study the properties of the SMM's objective function with explicit regard to the impact of broken ergodicity. Let us start with investigating the objective function's response surface. 
The aim is to check for discontinuities, (non-)smoothness and flat valleys, which create problems during the optimization as they impede finding global optima during the estimation.
Such plots provide further help in gaining an understanding of the mapping between model parameters and the objective function.

\begin{table}[h]
\begin{center}
\caption[Simulation Scenarios]{\textbf{Simulation Scenarios.} Allocation of limited budget of observations of $\Time \times \EnsembleSize \times \MCSize = \num{80 000 000}$ observations over different simulation dimensions.}
\tlabel{SimScenarios}
\vspace{1em}
\begin{tabular}{lcccccc}
\toprule
\multirow{2}{*}{\textbf{Scenario}}	& \multicolumn{2}{c}{Time only}	& \multicolumn{3}{c}{\MixNT mix} \\
\cmidrule(lr){2-3} \cmidrule(lr){4-7} 
& \Tshort	& \Tlong &  &  &  &  \\
& 	& (a)  & (b) & (c) & (d) & (e) \\ 
\toprule
Time Length	\Time								&	40k			&	400k	&	40k		& 20k		&	10k	& 40k \\
Ensemble Size \EnsembleSize			& 1				& 1			& 10		&	20		&	40	& 20 \\
Monte Carlo Runs \MCSize				&	200			& 200		& 200		& 200		& 200 & 200\\
\bottomrule
\end{tabular}
\end{center}
\end{table}
\FloatBarrier

\subsubsection{Response surface}
For this analysis, we will focus only on the most problematic model parameters in terms of identification power.
For the \ALWmodel, we have identified both herding parameters $a$ and $b$ as troublesome.
We run simulation scenarios summarized in \tref{SimScenarios} with a constant observation budget of \num{8000000} throughout all experiments, \ie one short time series realization $\Tshort = \num{40 000}$, a long time series with $\Tlong = \num{400 000}$ and over an ensemble with $\EnsembleSize = 10$ and $\Time = \num{40 000}$.
The parameter grid consists of 41 equidistant points between $\left[0.06,0.66\right]$ for parameter $a$ and between $\left[0.28,3.1\right]$ for parameter $b$.
The moment conditions are based on all 18 moment functions of \tref{MomentFunctions}. \Fref{3Dplots} shows the corresponding results for three different sample simulation runs using different random seeds. The solid lines mark the true value of the model parameters $\left(a,b\right)=\left(0.0003,0.0014\right)$.
The intersection of the two lines build the theoretical optimum of the objective function.
Note that we plot the response surface of the inverse of the objective function to better visualize the global optimum.
Ideally, the highest peak on the response surface coincides with the intersection of the true parameter values.

\begin{figure}[pb]
    \begin{subfigure}{\textwidth}
	    \centering
	    \includegraphics[width=.79\textwidth]{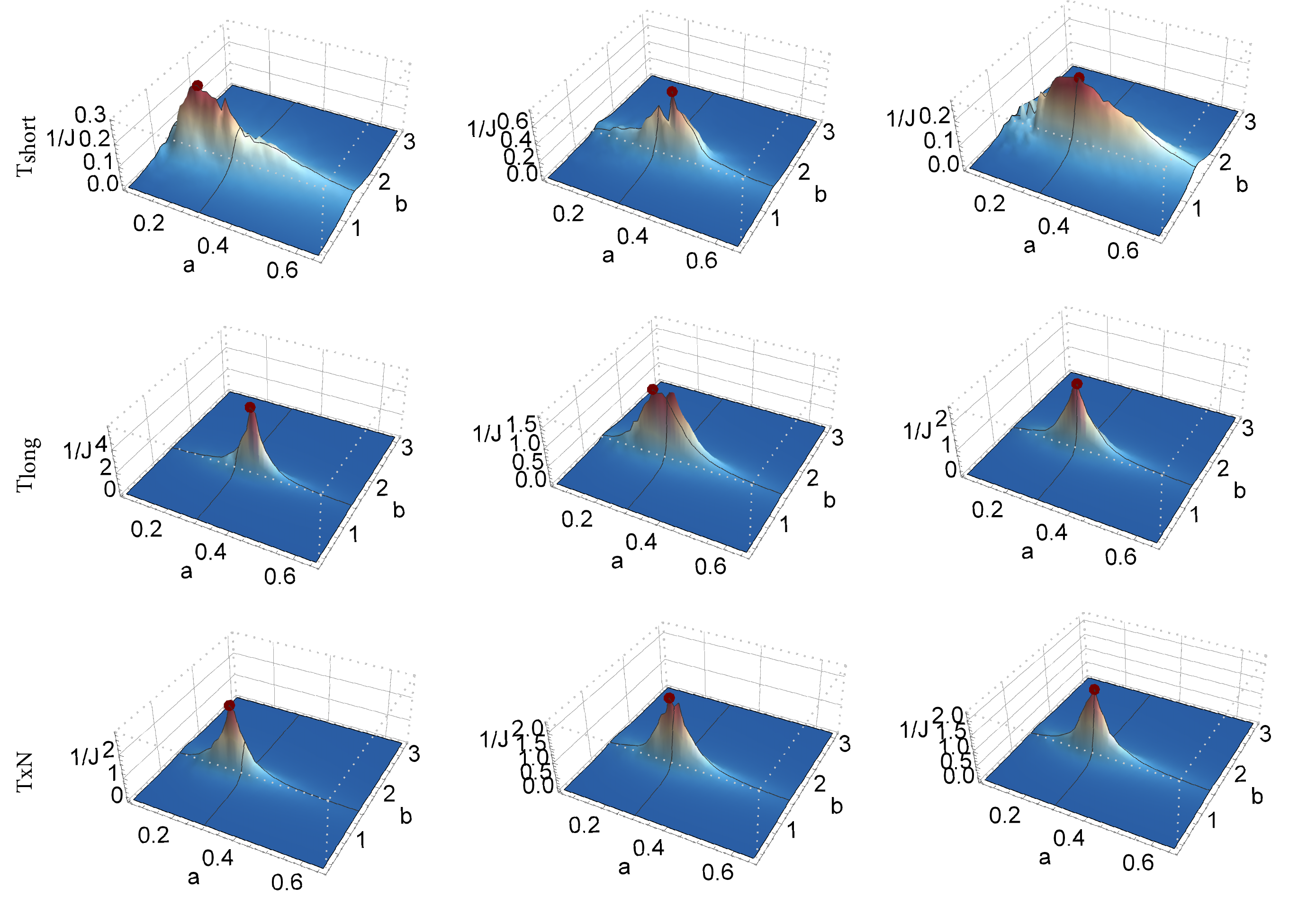}
	    \caption{\ALWmodel}
	    \flabel{3DplotALW}
    \end{subfigure}
    \vspace{1em}
    \begin{subfigure}{\textwidth}
	    \centering
      \includegraphics[width=.79\textwidth]{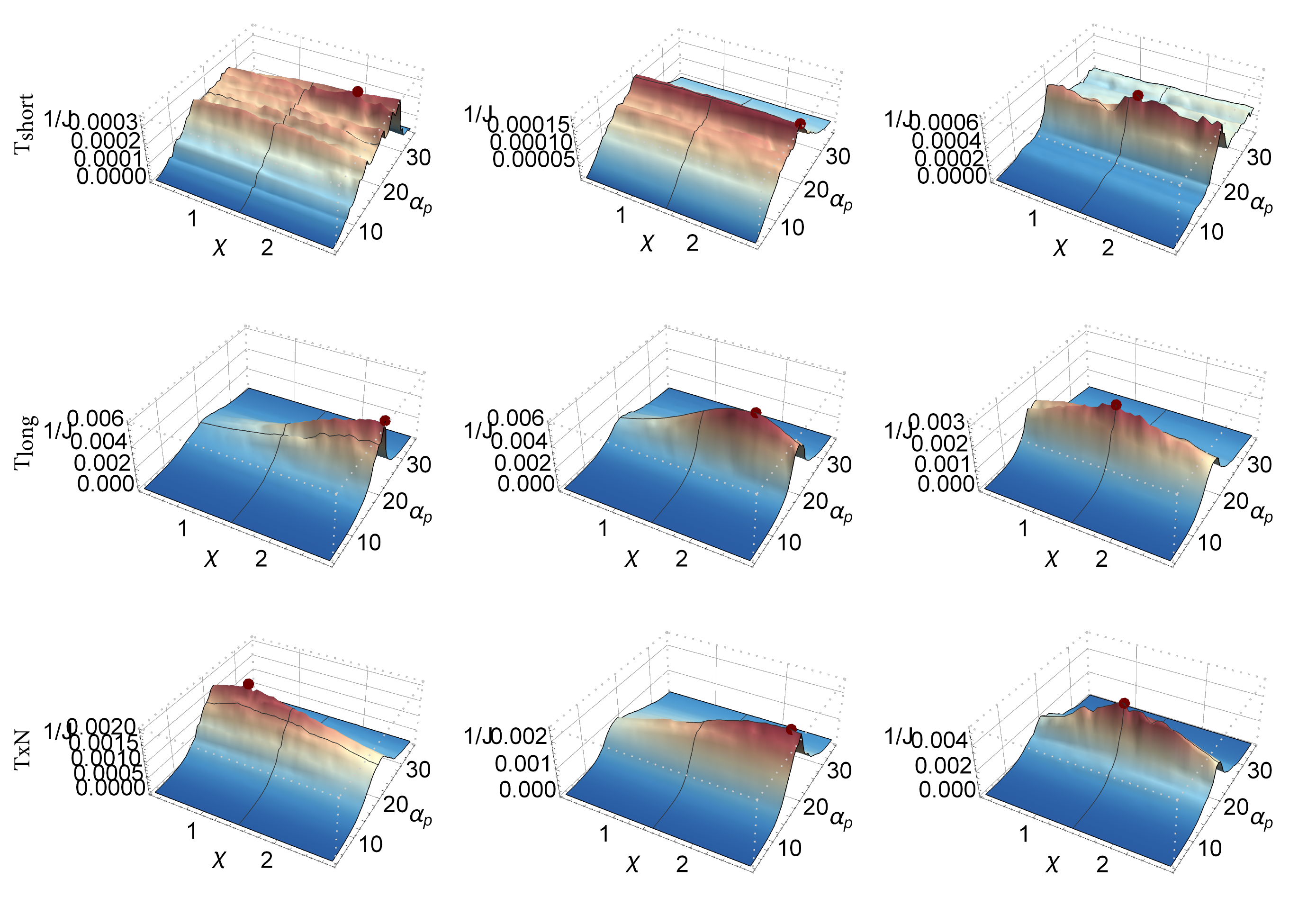}
      \caption{\FWmodel}
	    \flabel{3DplotFW}
    \end{subfigure}
    \caption{\textbf{3D surface plots for critical model parameters.}
    Contour plots of the inverse objective function $\ObjF^{-1}$ for varying values for parameter pairs ($a, b$) for the \ALWmodel in \fref{3DplotALW} and ($\chi, \alpha_p$) for the \FWmodel in Panel \fref{3DplotFW}.
    Sample simulations are run over different settings: $\Tshort = \num{40 000}$, $\Tlong = \num{400 000}$ and \MixNT with $\Time = \num{40 000}$ and $\EnsembleSize = 10$.
    Black lines indicate the (pseudo-)true parameter values for selected model parameters and red dots denote the highest peak of the inverse of the objective function's response surface.}
   \flabel{3Dplots}
\end{figure}

The response surface of the objective function is clearly non-smooth and contains multiple local optima.
Depending on the choice of random seeds, the fitness of the objective function differs greatly.
While parameter $b$ is correctly identified in all three simulation samples, estimates of parameter $a$ happen to be far (left plot) to slightly off (middle plot) compared with the true value of $a=0.0003$.
One might be even lucky enough to identify both parameters correctly (right plot).

Increasing the number of observations from $\Time = \num{40 000}$ to $\Time = \num{400 000}$, we find a smoother surface as well as less variations as expected.
However, even for this higher sample size, local optima arise (see the middle plot) depending again on the choice (or luck) of random seeds.
Comparing this with an ensemble of $\EnsembleSize = 10$ simulation runs for $\Time = \num{40 000}$, we qualitatively observe the same patterns.

During a pre-exploratory experiment for the \FWmodel, we identified the most problematic parameter pairs ($\chi, \alpha_0$) and ($\chi$,$\alpha_p$). The latter pair is visualized in \fref{3Dplots}.\footnote{Graphical results for the objective function's response surface of the \FWmodel are almost identical for parameter pairs ($\chi$,$\alpha_0$) and ($\chi$,$\alpha_p$).}
In the following, we set the (pseudo-)true values $\left(\chi,\alpha_p\right)=\left(1.5,19.671\right)$ and grid points result from 41 equidistant parameter variations over $\left[0.3,2.7\right]$ for parameter $\chi$ and over $\left[3.9342,3.9342\right]$ for parameter $\alpha_p$.
Let us first note that we find qualitatively the same results as for the \ALWmodel.
Yet, the contour plots look rather different for the \FWmodel.
We observe that the surface is less elevated with peaks at much lower levels than we see for the \ALWmodel.
Especially for the \Tshort-setting, we find very flat hilltops for parameter $\chi$ and even flat plateaus for both parameters.
Increasing the number of observations helps to smoothen the surface.
However, identification issues leading to biases in the estimates might be more pronounced here which is most probably due to the higher dimensionality and complexity of the \FWmodel.

While these contour plots of the objective function reveal a partially non-smooth surface, they allow no assessment of the estimator's further distributional properties. To do so, we need to include the third dimension of our simulation cube shown in \fref{MonteCarloCube} which is referred to the Monte Carlo repetitions.
Hence, we run the simulations repeatedly over $\MCSize = 200$ and present graphical results in \Fref{BoxplotHisto}. The left panel shows boxplots for the three different simulation settings.
First, we notice that for the \ALWmodel the estimator is consistent and overall able to detect the true value of parameter $a$ on average.
Unsurprisingly, variations are comparatively larger for the \Tshort-setting.
Regarding the other two settings, variations differ only slightly with almost unbiased estimates for the \Tlong-setting and a small downward bias in the mix-setting.
The paired histogram plot on the right panel in \fref{BoxplotHisto} confirms this observation.
While for the \Tlong-simulation the histogram is quite symmetric, the histogram for the mix-simulation is skewed towards lower values of $a$.

For the \FWmodel results differ slightly. We find small biases of the estimates for the \Tshort and \Tlong-cases while the \MixNT-setting is able to perfectly identify the true value of parameter $\chi$ on average. This suggest that the ensemble setting might lead to more consistent and efficient estimates. However, we are cautious to not overinterpret our results here given that we have included only two varying parameters while keeping the rest fixed.
\begin{figure}[h]
\begin{center}
  \subcaptionbox{\ALWmodel.\flabel{BoxplotHistoALW}}{\includegraphics[width=.85\textwidth]{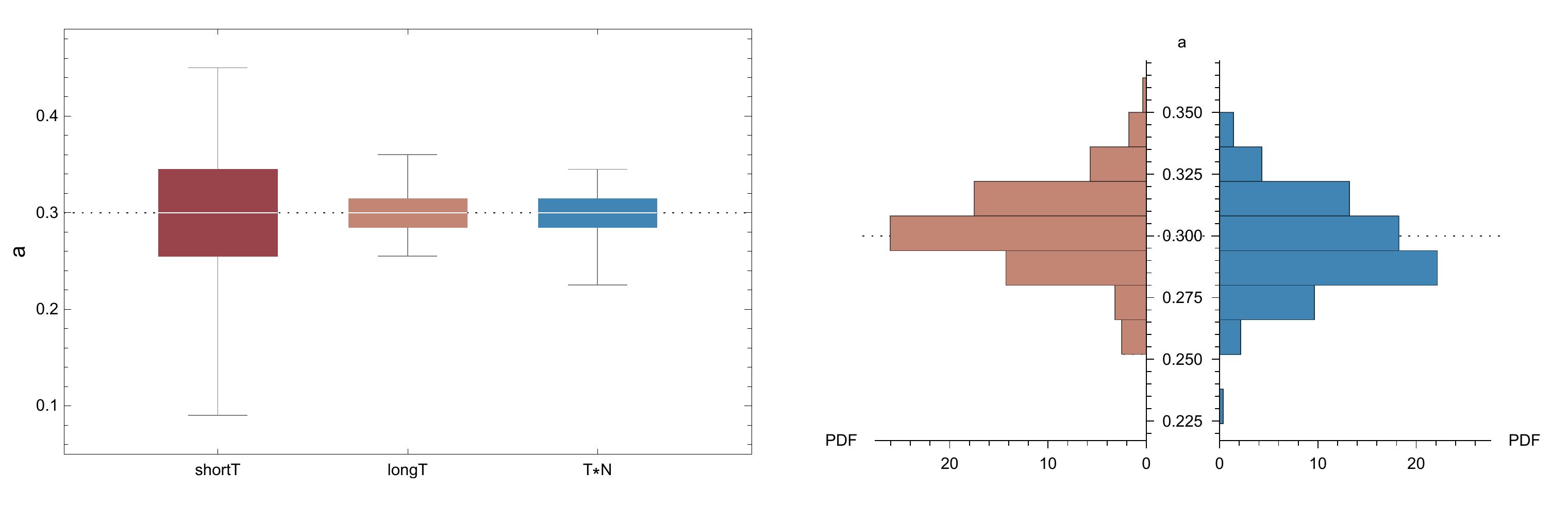}}\\
	\subcaptionbox{\FWmodel.\flabel{BoxplotHistoFW}}{\includegraphics[width=.85\textwidth]{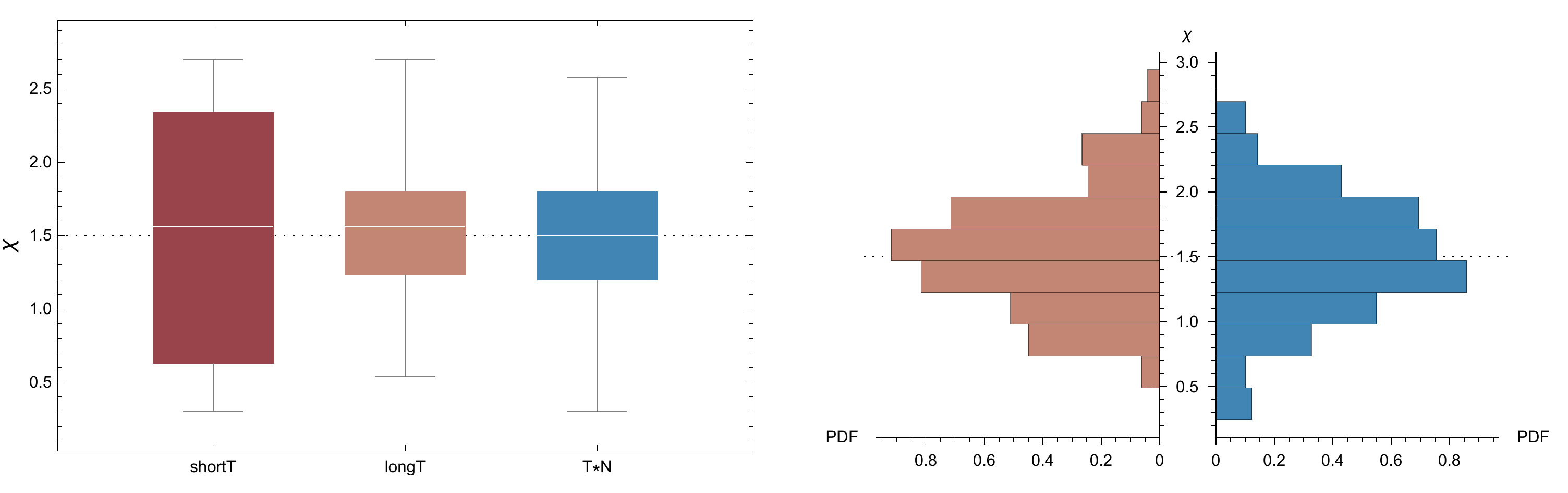}}
\end{center}
\caption{\textbf{Box plots and paired histograms for critical model parameters.}
Upper panel \fref{BoxplotHistoALW} for the \ALWmodel and lower panel \fref{BoxplotHistoFW} for the \FWmodel.
Left figures show box plots of the estimates for parameter $a$ for the \ALWmodel and parameter $\chi$ for the \FWmodel.
The settings are the following: $\Tshort = \num{40 000}$, $\Tlong = \num{400 000}$ and \MixNT with $\Time = \num{40 000}$, $\EnsembleSize = 10$, whereby the latter two have the same budget of observations.
Simulations are run over $\MCSize=200$ Monte Carlo repetitions.}
\flabel{BoxplotHisto}
\end{figure}

\subsubsection{Numerical estimation experiment}
This brings us directly to the next experiment where we include all model parameters for estimation.
In the following, we will run a small estimation exercise over a Sobol-sequenced sample of $\num{2000}$ parameter combinations.
The included parameter space is given by $\pm25\%$ of the true values which results in the following parameter ranges: $a \in \left[0.225,0.375\right]$, $b \in \left[1.05,1.75\right]$, $\sigma_f \in \left[22.5,37.5\right]$ for the \ALWmodel and $\phi \in \left[0.09,0.15\right]$, $\chi \in \left[1.125,1.875 \right]$, $\alpha_0 \in \left[-0.252,-0.42\right]$, $\alpha_n \in \left[1.379,2.299\right]$, $\alpha_p \in \left[14.753,24.589\right]$, $\SigmaF \in \left[0.531,0.885\right]$, $\SigmaC \in \left[1.61,2.684\right]$ for the \FWmodel.\footnote{The defined parameter ranges provide broad parameter variations while keeping execution times of the simulation runs feasible.} 
The number of Monte Carlo runs is set again to $\MCSize = 200$.
Since we are interested in the degree of uncertainty that solely comes from broken ergodicity, we abstain from running a full estimation exercise given that the choice of the optimization algorithm would highly influence and distort our results.\footnote{\textcite{WinkerMaringer2009} provide a study of the joint convergence of an estimator and a heuristic optimization algorithm. They compute the necessary number of Monte Carlo repetitions of the optimization routine to derive robust estimates.}
We report graphical results in \fref{BoxplotEsti} and the corresponding mean estimates together with their standard deviations and root-mean squared errors in \tref{EstiResults}.
Looking at \fref{BoxplotEsti}, the left vertical axis shows the value of the objective function \ObjF. Settings (a) and (b) result in the lowest values of \ObjF.
For the \ALWmodel, it seems that both (a) and (b) are compatible in terms of minimized objective function.
Mean sample estimates in \tref{EstiResults} confirm these findings.
Decreasing the length of \Time while increasing the ensemble size \EnsembleSize raises the level of uncertainty in the estimates, pushing mean estimates slowly away from their true values.
Accordingly, standard deviations and root-mean squared errors tend to increase, too.

Regarding the \FWmodel and its comparatively higher dimensionality, results are qualitatively similar yet the loss in efficiency for increasing ensemble sizes \EnsembleSize is more pronounced here.
This loss in efficiency becomes particularly obvious for setting (d) with $\Time = \num{10 000}$ and $\EnsembleSize = 40$. In fact, the number of observations per realized time series is too short here resulting in uninformative samples.

Summing up, we find different values of the objective function \ObjF for different allocations of the total budget of observations. While the differences might be negligible for longer time lengths \Time, they significantly deteriorate estimates for shorter time lengths due to broken ergodicity.
Recall the ergodic theorem in \eref{ergodicInequality} which is effective here.
\begin{figure}[htp]
	\subcaptionbox{\ALWmodel.\flabel{BoxplotEstiALW}}{\includegraphics[width=.5\textwidth]{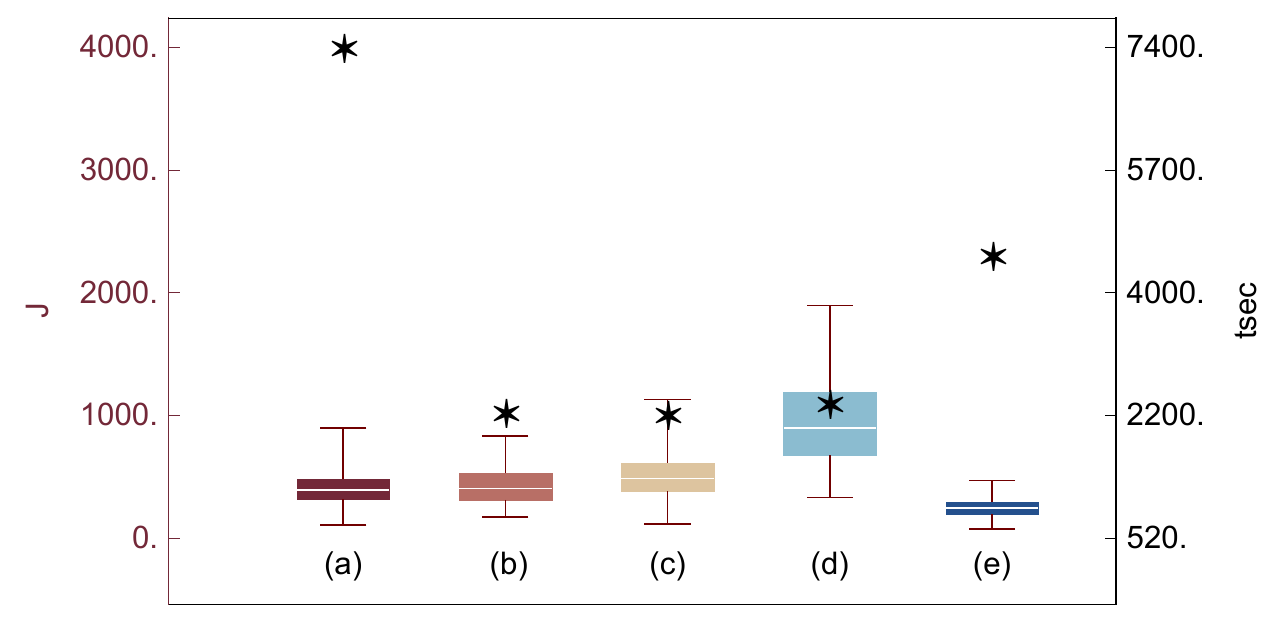}}
	\subcaptionbox{\FWmodel.\flabel{BoxplotEstiFW}}{\includegraphics[width=.5\textwidth]{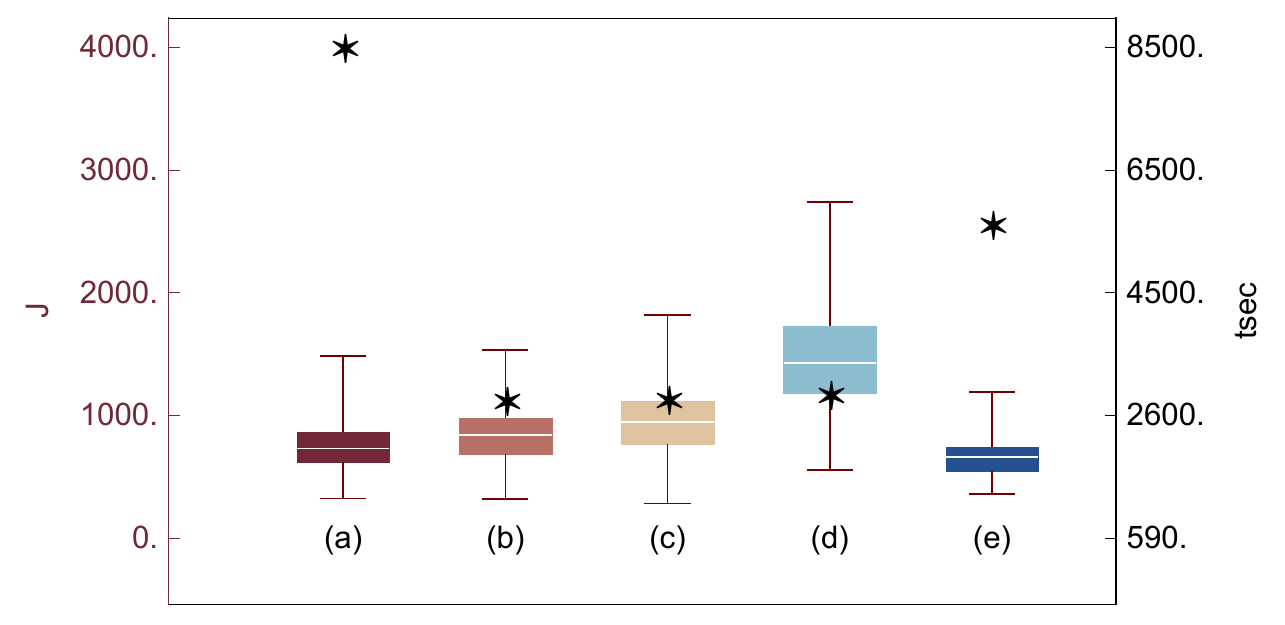}}
	\caption{\textbf{Box plots and timing results of estimation exercise.}
	Box plots of objective function \ObjF (left vertical axes) together with star points representing the required computational time to run the simulations (right vertical axes).
	Left panel \fref{BoxplotEstiALW} shows results for the \ALWmodel.
	Right panel \fref{BoxplotEstiFW} shows results for the \FWmodel.
	The scenarios allocate differently the total budget of \num{400 000} observations in the following ways:
	(a) $\Time = \num{400 000}$, $\EnsembleSize = 1$;
	(b) $\Time = \num{40 000}$, $\EnsembleSize = 10$;
	(c) $\Time = \num{20 000}$, $\EnsembleSize = 20$;
	(d) $\Time = \num{10 000}$, $\EnsembleSize = 40$ and
	(e) $\Time = \num{40 000}$, $\EnsembleSize = 20$.
	Simulations are run over $\MCSize=200$ Monte Carlo repetitions.}
	\flabel{BoxplotEsti}
\end{figure}

However, investigating only the finite sample properties of the parameter estimates neglects the computational resources needed to run these simulations.
Therefore, we add the execution times as black stars, belonging to the right vertical axis, to \fref{BoxplotEsti}.
Setting (a) is by far the most computationally demanding simulation.
The three \MixNT-settings (b), (c) and (d) have similar execution times which reduce the benchmark setting (a) by a factor of three for both the \ALWmodel and \FWmodel.
The reason is that single simulations over \Tlong are memory intense and inefficient.
In fact, repeated simulations over different random seeds are so called \textit{embarrassingly parallel} tasks and can, thus, be easily distributed over multiple processing units which is done here.\footnote{All numerical simulations in this paper have been performed using Julia 1.0.5 on a standard desktop computer with an eight core 3.00 Ghz AMD Ryzen 7 1700 CPU and 32GB of memory. We made sure to run all simulation settings most efficiently to keep the different scenarios comparable.}
We finally run a fifth scenario (e) for which we take the best performing \MixNT-setting (b) with $\Time = \num{40 000}$ but double the amount of ensemble size $\EnsembleSize = 20$. This setting is able to easily outperform all other settings, even scenario (a) while taking only two third of setting (a)'s computational time. We find this to be a consistent pattern for both FABMs considered here.

With this analysis, we are not only aiming for an adequate mixture of \MixNT from the perspective of (broken) ergodicity but also considering the associated computational costs.
We conclude that for a limited budget of computational resources, using a mix of \MixNT is favourable over a single long simulation run with the same number of observations.

\begin{table}[h]
\caption{\textbf{Estimation results.}
The table lists mean estimates, their standard deviations and root-mean squared errors of the estimated parameters for the scenarios (a)-(e) given in \tref{SimScenarios}. We also report \ObjF-values in the last column of the \ALWmodel and \FWmodel, respectively. Results are based on $\MCSize=200$ repetitions.}
\tlabel{EstiResults}
\scriptsize
\begin{tabular}{lrrrrrrrrrrrrr}
\addlinespace
\toprule
    &        & \multicolumn{4}{c}{\ALWmodel}  & \multicolumn{8}{c}{\FWmodel} \\
\cmidrule(lr){3-6} \cmidrule(lr){7-14}
    &        & $a$    & $b$    & \SigmaF     & \ObjF	& $\phi$	& $\chi$	& $\alpha_0$	& $\alpha_n$	& $\alpha_p$	& \SigmaF	& \SigmaC	& \ObjF\\
\cmidrule(lr){3-14}    
\addlinespace
    & \paramVecTrue & 0.3   & 1.4   & 30     & 0      & 0.12  & 1.5   & -0.336 & 1.839 & 19.671 & 0.708 & 2.147 & 0       \\
\cmidrule(lr){2-14}
\addlinespace
(a) & mean   & 0.298 & 1.4   & 30.003 & 400.09 & 0.123 & 1.801 & -0.287 & 1.871 & 19.153 & 0.702 & 2.19  & 747.07  \\
    & STD    & 0.03  & 0.019 & 0.219  & 125.26 & 0.016 & 0.108 & 0.024  & 0.135 & 1.253  & 0.009 & 0.11  & 193.05  \\
    & RMSE   & 0.03  & 0.019 & 0.218  & 0      & 0.016 & 0.32  & 0.055  & 0.138 & 1.353  & 0.011 & 0.118 & 0       \\
\addlinespace
\midrule
\addlinespace    
(b) & mean   & 0.292 & 1.405 & 30.017 & 423.84 & 0.123 & 1.785 & -0.284 & 1.88  & 18.905 & 0.702 & 2.194 & 829.72  \\
    & STD    & 0.026 & 0.022 & 0.207  & 149.98 & 0.015 & 0.147 & 0.019  & 0.131 & 0.984  & 0.01  & 0.119 & 222.86  \\
    & RMSE   & 0.027 & 0.022 & 0.208  & 0      & 0.015 & 0.321 & 0.056  & 0.137 & 1.245  & 0.012 & 0.128 & 0       \\
\addlinespace
\midrule
\addlinespace    
(c) & mean   & 0.288 & 1.412 & 30.086 & 504.92 & 0.121 & 1.696 & -0.296 & 1.925 & 18.564 & 0.707 & 2.153 & 963.66  \\
    & STD    & 0.03  & 0.021 & 0.258  & 179.94 & 0.017 & 0.233 & 0.032  & 0.14  & 1.184  & 0.013 & 0.126 & 262.44  \\
    & RMSE   & 0.033 & 0.024 & 0.272  & 0      & 0.017 & 0.304 & 0.051  & 0.164 & 1.619  & 0.013 & 0.126 & 0       \\
\addlinespace
\midrule
\addlinespace    
(d) & mean   & 0.287 & 1.426 & 30.111 & 948.87 & 0.123 & 1.579 & -0.306 & 1.982 & 18.212 & 0.714 & 2.109 & 1481.87 \\
    & STD    & 0.028 & 0.024 & 0.282  & 344.38 & 0.016 & 0.249 & 0.038  & 0.116 & 1.111  & 0.013 & 0.112 & 386     \\
    & RMSE   & 0.031 & 0.035 & 0.302  & 0      & 0.016 & 0.261 & 0.048  & 0.184 & 1.832  & 0.015 & 0.118 & 0       \\
\addlinespace
\midrule
\addlinespace    
(e) & mean   & 0.29  & 1.403 & 30.033 & 244.35 & 0.128 & 1.832 & -0.277 & 1.897 & 18.634 & 0.702 & 2.179 & 622.62  \\
    & STD    & 0.019 & 0.013 & 0.162  & 70.73  & 0.009 & 0.084 & 0.01   & 0.068 & 0.49   & 0.005 & 0.065 & 156.9   \\
    & RMSE   & 0.022 & 0.013 & 0.165  & 0      & 0.012 & 0.342 & 0.06   & 0.089 & 1.147  & 0.008 & 0.073 & 0    \\
\bottomrule    
\end{tabular}
\end{table}

\FloatBarrier

\subsubsection{Robustness of results}
\begin{figure}[h]
\centering
	\subcaptionbox{\ALWmodel.\flabel{MomSensiALW}}{\includegraphics[height=4.5cm]{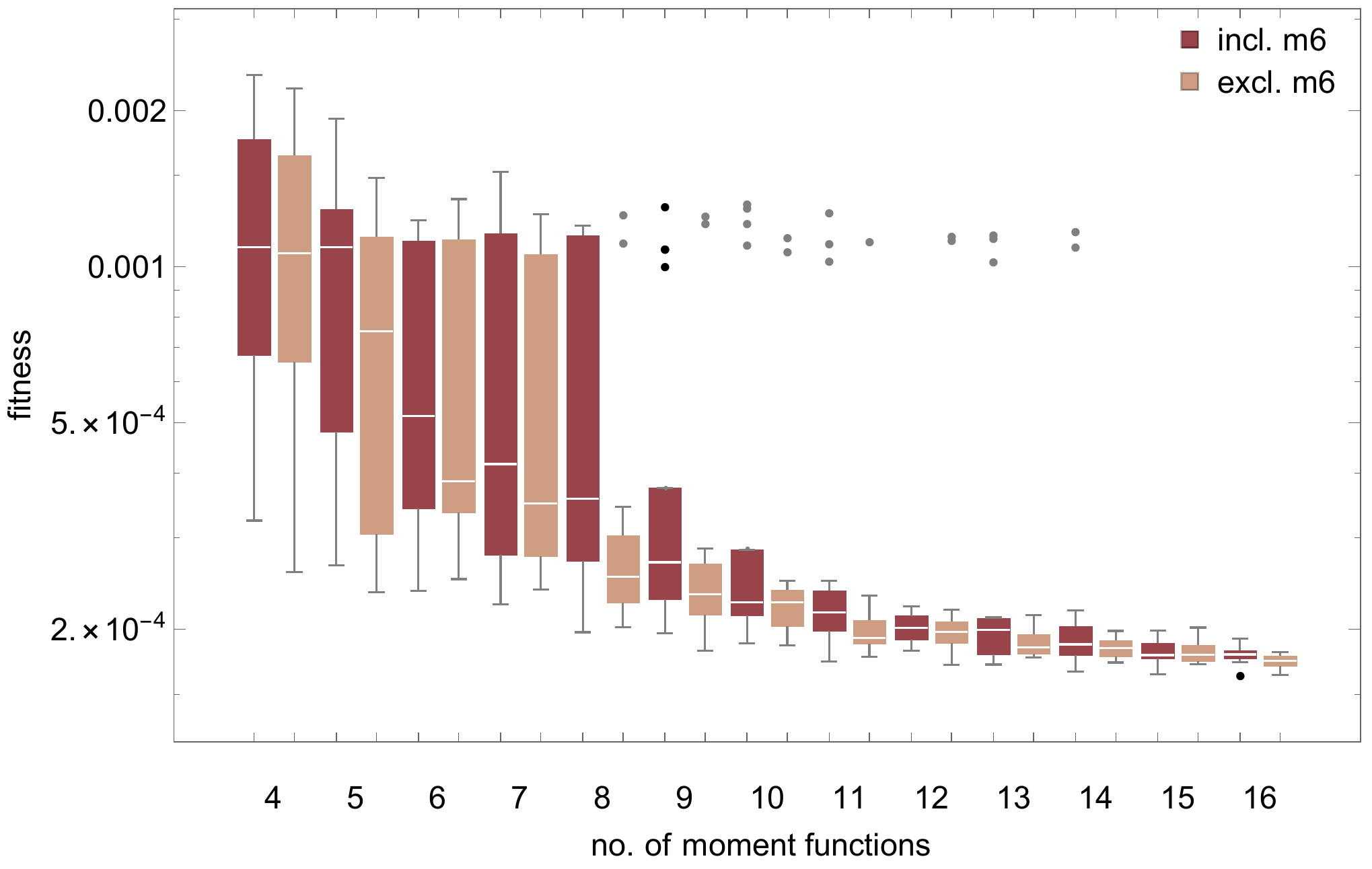}}
	\subcaptionbox{\FWmodel.\flabel{MomSensiFW}}{\includegraphics[height=4.5cm]{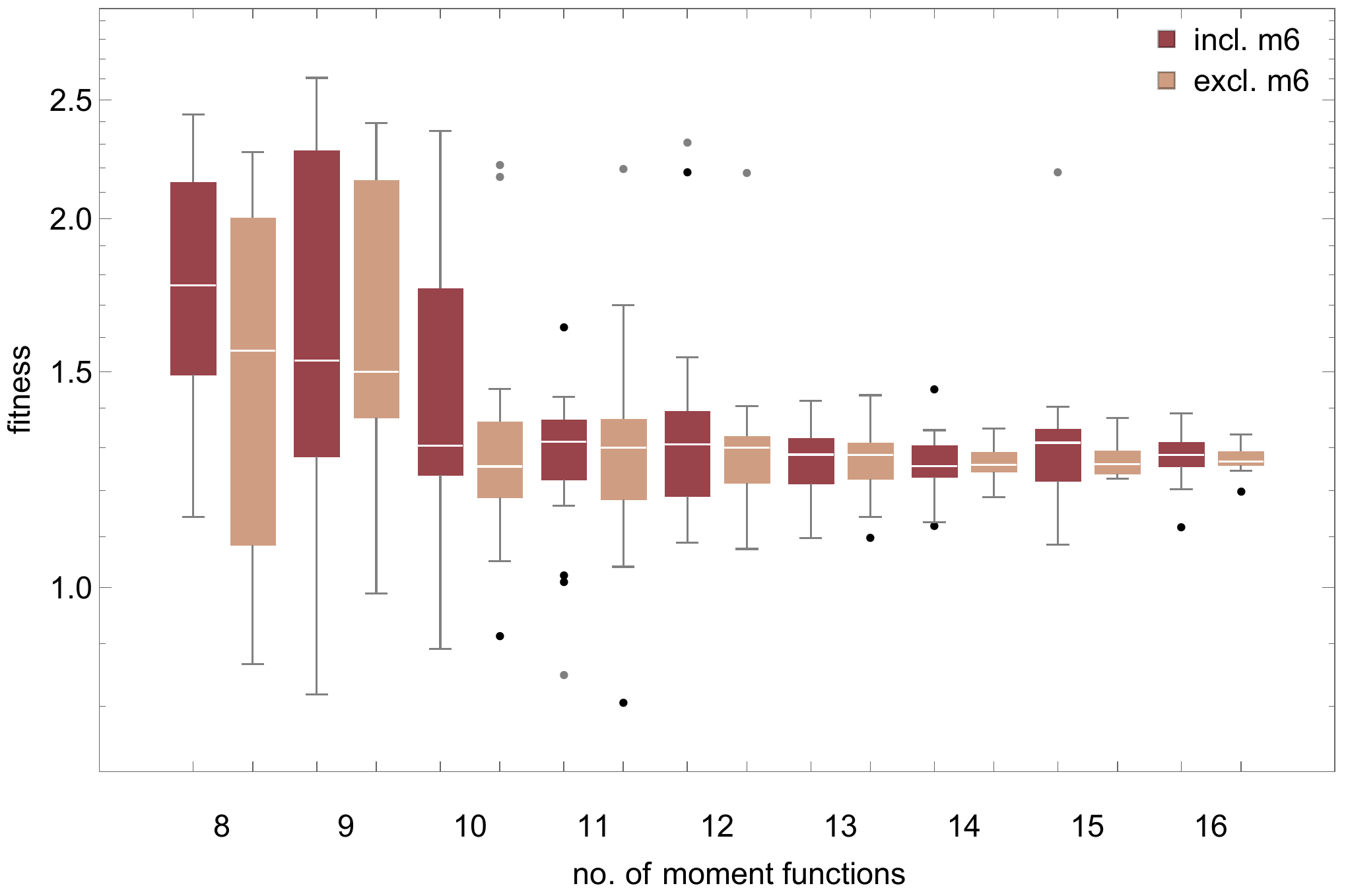}}
	\caption{\textbf{Sensitivity analysis for varying number of moment functions.}
	Boxplots show improved fitness for increased moment vector size with fitness defined as the euclidean distance between the (pseudo-)true and estimated parameter vector.
	We consider two different settings for the sampling of moment functions: dark red boxplots including moment function $m_6$ an orange boxplots excluding $m_6$.
	Left panel \fref{MomSensiALW} shows results for the \ALWmodel.
	Right panel \fref{MomSensiFW} shows results for the \FWmodel.
	Note that we use a logarithmic scale for the vertical axis to better visualize the fitness.}
	\flabel{MomSensi}
\end{figure}

How robust are these numerical results especially regarding variations of the moment function set?
In the following, we try to answer this question by running sensitivity analyses for varying sizes \MomentsVecSize of the moment vector \momVec.
For this, we randomly sample 17 different moment sets per number of moment vector size and evaluate them for the numerical estimation data of setting (a).
Fitness is defined as the euclidean distance between the (pseudo-)true and estimated parameter vector.
We decide to investigate two different scenarios here.
For the first scenario we sample from the complete range of 18 moment functions as listed in \tref{MomentFunctions}.
For the second scenario we exclude $m_6$ known as the autocorrelation of raw returns at lag 1.
The reason is that we have seen for example in \fref{MomentParameterSensitivityALW} problematic to non-convergent behaviour for moment function $m_6$.
We suspect improved results in terms of fitness given that $m_6$ offers very limited identification power.
We present graphical results in \fref{MomSensi}.
The two figures show boxplots for the \ALWmodel on the left and the \FWmodel on the right.
Unsurprisingly, results improve for an increasing number of included moment functions.
More interestingly and rather unexpectedly, the improvement does not happen linearly.
While variations in the fitness are quite large for smaller moment vector sizes, we observe a sudden drop for moment vector sizes of nine (\ALWmodel) and ten (\FWmodel), respectively.
After that drop results are pretty robust for further increasing number of moments.
This is especially true for the \FWmodel while we see slight improvements for the \ALWmodel.
Comparing both scenarios (with and without $m_6$, graphically represented by dark red and orange) we find consistently better fitness for moment sample sets excluding the autocorrelation of raw returns at lag 1.
Additionally, estimates related to this scenario tend to produce less outliers (see the grey and black dots).

This suggest that re-evaluating our estimation results from \tref{EstiResults} for an adjusted moment vector should lead to significantly better results.
Therefore, we take our previous results as presented in \tref{EstiResults} and recompute them for a reduced moment vector excluding $m_6$.
We relabel these results $J_{18}$ since they have been actually estimated for a complete set of 18 moments.
Now we can compare them with the $J_{17}$ results that we estimated based on 17 moment functions.
We report the differences between $J_{18}$ and $J_{17}$ values for all five settings in \tref{RobustnessJ}.
As we anticipated, we find clearly improved \ObjF-values as a common and consistent pattern.
Furthermore, one-tailed $t$-tests reveal statistical significance for all results except for setting (d).

\begin{table}[h]
\centering
\caption{\textbf{\ObjF results for adjusted moment vector.}
For the scenarios $(a)-(e)$ given in \tref{SimScenarios}.
Significance levels of a one-tailed $t$-test are represented by \oneS, \twoS and \threeS for 10\%, 5\% and 1\%, respectively.}
\tlabel{RobustnessJ}
\begin{tabular}{llllll}
\addlinespace
\toprule
    & \multicolumn{5}{c}{$\ObjF_{18} - \ObjF_{17}$}    \\
        \cmidrule(lr){2-6}    
\addlinespace
    & (a)     & (b)   & (c)   & (d)   & (e)   \\
    \cmidrule(lr){1-6}    
\addlinespace
\ALWmodel & 23.84\threeS	& 26.24\threeS	& 23.15\twoS	& 23.25	& 12.17\threeS \\
\FWmodel  & 27.42\twoS		& 34.02\twoS		& 29.18\oneS	& 24.8	& 17.99\twoS\\
\bottomrule
\end{tabular}
\end{table}
\FloatBarrier

\section{Conclusion}\seclabel{Conclusion}

(F)ABMs are often conceived as black boxes lacking a sound and well-behaved mathematical model.
This has caused considerable resistance in the (economics) community to accept computer simulations as proper research methods.
The development of estimation and validation tools that are particularly suited for the special properties of most (F)ABMs is a very active field of research.
This paper aims to contribute to this line of research by explicitly considering the uncertainty coming from broken ergodicity.

We systematically study the properties and convergence behaviour of a SMM estimator and its individual moment functions.
We have seen that assuring a one-to-one mapping between model output in terms of moment functions and model parameters is not trivial.
Most moment functions are indeed non-linear and non-monotone.
This can lead to biased estimations.
More important is however the relative responsiveness of moment functions towards changes in the model parameters, \ie how strongly they react.
Parameters for which we find weak responsiveness of moment functions tend to be the most troublesome during estimation.
We have further shown that not all moment functions which are a considered to be a popular choice for the estimation of univariate asset pricing models are actually suited.
As it turns out, the convergence of the autocorrelation of raw returns at lag 1 is highly distorted making it an unfavourable candidate statistic.
We show that estimation results improve significantly when leaving out this summary statistic.
For all other moment functions we find robust results for the moment matching fitness.
Therefore, our analysis confirms our choice of moment functions for (univariate) asset price models and thereby offers a response towards the arbitrariness critique of SMM.


Estimation methods like SMM are perfect tools for perfect models.
Living in an ever-changing world, it should be no design fault to create models that possess non-ergodic properties. 
Even if regularity conditions are not fully met, models and tools may still be useful.
Being aware of broken ergodicity is the key here.
We further suggest researchers to not blindly follow previous studies based on other models.
In fact, every (F)ABM behaves differently as we have seen here.
Therefore, we run a bunch of Monte Carlo analyses to learn about the FABMs' sensitivity and responsiveness towards changes in model parameters and simulation settings.
We have shown that for a given computational time budget an adequate mixture of ensemble size \EnsembleSize and time length \Time is better suited than the same number of observations in one realization.
We can conclude that due to careful selection of moment functions and consideration of ergodicity, we can improve the objective function's potential to correctly identify model parameters for a limited budget of computational resources.

So far, the energy demand of simulation models seems to play a minor role.
Access to institute-owned high performance computing clusters and the recent global increase in cloud computing may have made the necessity for energy-efficient simulations obsolete.
Yet, the opposite is true given the looming threat of climate change and a likely exponential increase in demand for simulations like FABMs especially for policy purposes in the near future.
A holistic approach to fight global warming will include many different aspects of our lives.
This also includes an efficient and mindful use of resources when it comes to research.
Our work provides steps into this direction.

We have limited our focus in this paper on the use of an SMM estimator.
The scale of our work can be extended to including other estimation methods as well.
We specifically aim to apply our approach to likelihood-based methods.
Since computing the likelihood function is expensive in terms of computational costs, applying our insights might improve cost efficiency there too.


\newpage
\printbibliography
\newpage
\appendix
\section{Sample simulation runs of model dynamics}\label{appendix:AppendixSampleSimulationRuns}

\begin{figure}[htp]
\begin{center}
  \subcaptionbox{\ALWmodel.\flabel{dynamicsALW}}{\includegraphics[width=.55\textwidth]{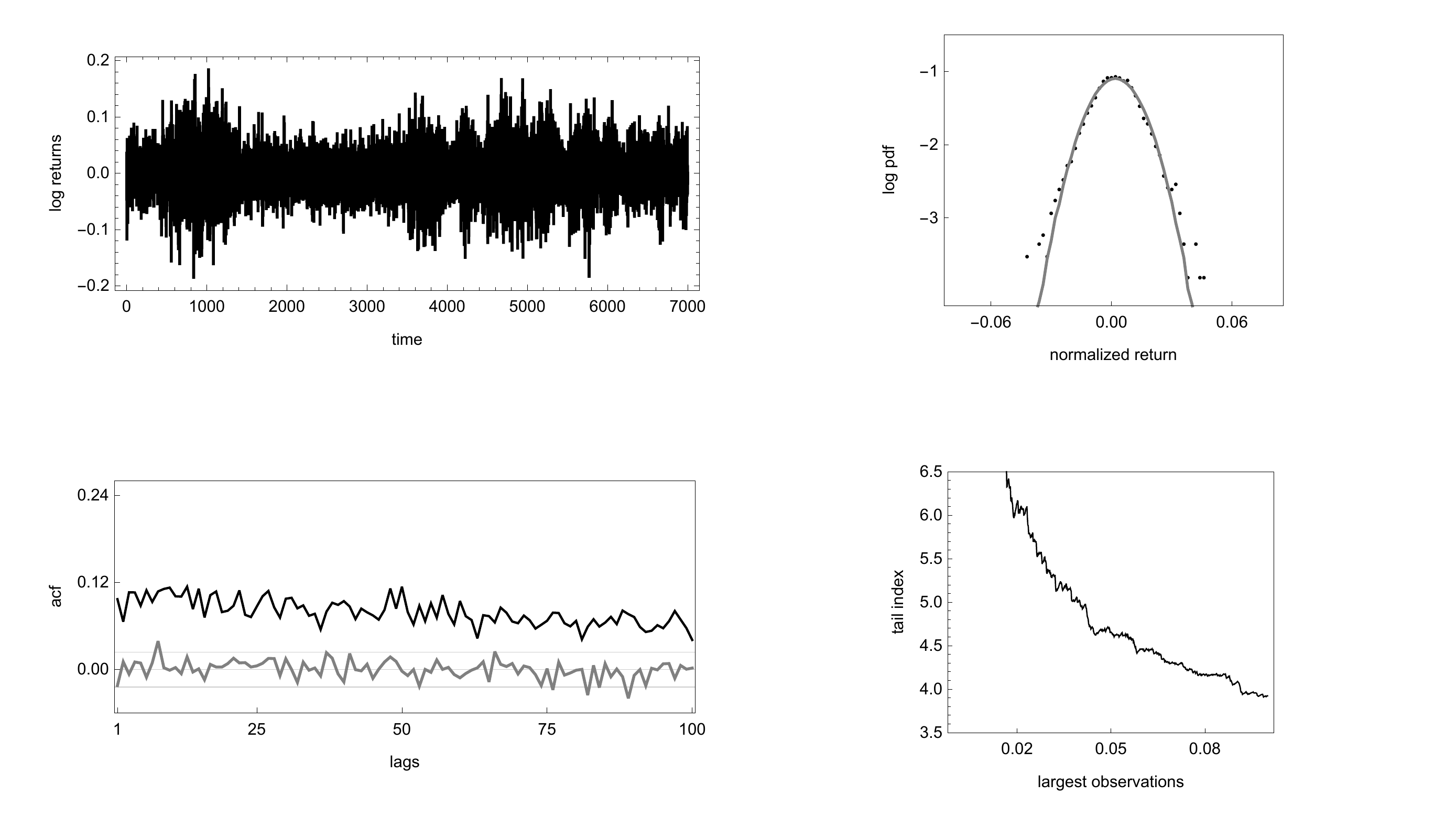}}\\
	\subcaptionbox{\FWmodel.\flabel{dynamicsFW}}{\includegraphics[width=.55\textwidth]{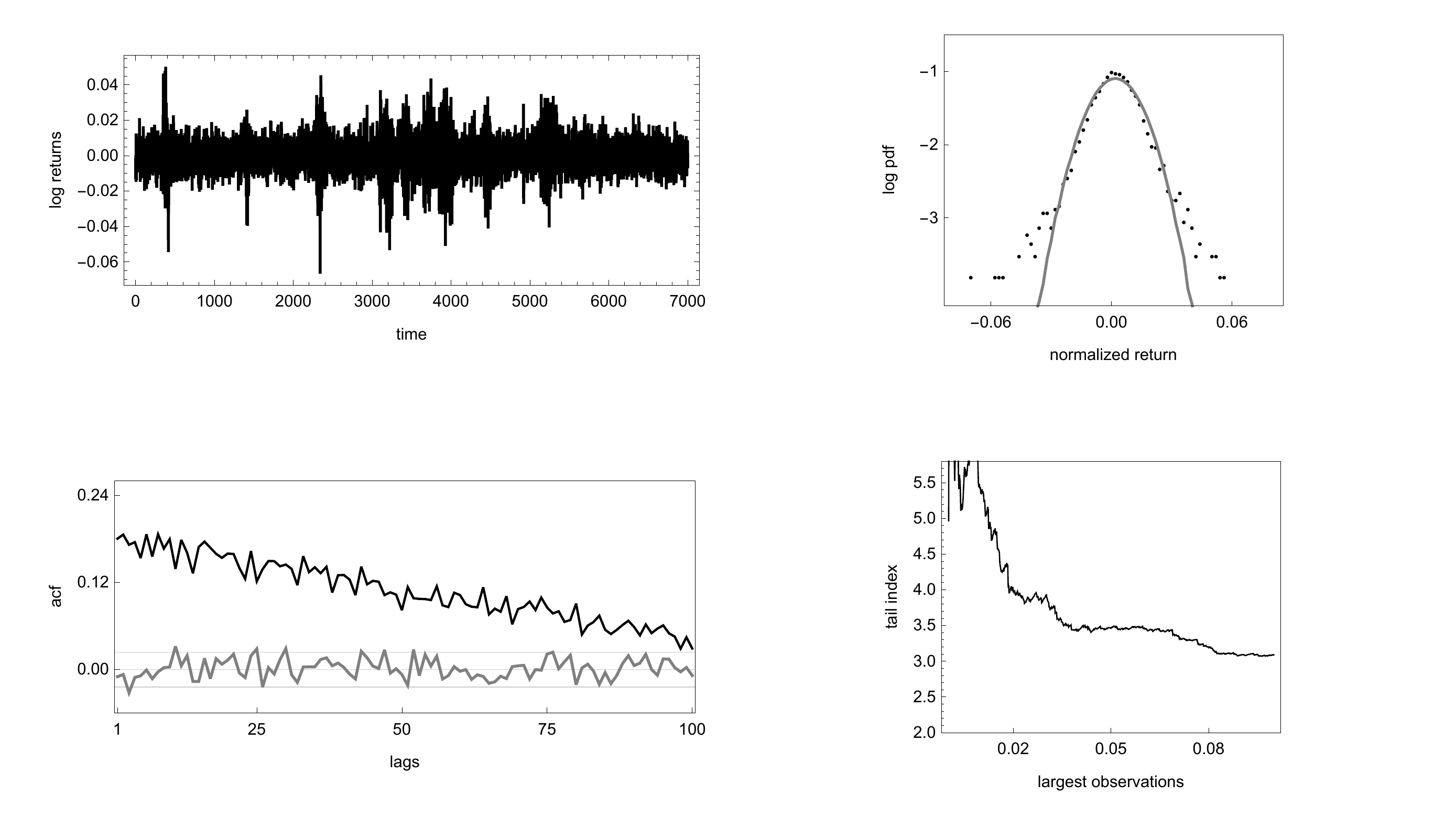}}
\end{center}
\caption{\textbf{Model dynamics of sample simulation runs.}
Upper panel \fref{dynamicsALW} shows dynamics for the \ALWmodel using the following parameter set: $a=0.3$, $b=1.4$, $\sigma_f=30$. The lower panel \fref{dynamicsFW} reveals dynamics for the \FWmodel using the following parameter set: $\phi=0.12$, $\chi=1.5$, $\alpha_0=-0.336$, $\alpha_n=1.839$, $\alpha_p=19.671$, $\SigmaF=0.708$, $\SigmaC=2.147$.
The panels show from top left clockwise to bottom left the evolution of log returns, the log probability density functions of normalized returns (black) and standard normally distributed returns (gray), the Hill tail index estimator and the autocorrelation functions of raw (gray) and absolute returns (black), respectively.}
\flabel{ModelDynamics}
\end{figure}
\FloatBarrier


\end{document}